\documentclass[authoryear,preprint,review,12pt]{elsarticle}

\newcommand{\beq}{\begin{equation}}
\newcommand{\eeq}{\end{equation}}

   \usepackage{epsfig}

\usepackage{amssymb}

\usepackage{natbib}

\begin{document}

\begin{frontmatter}



\author[rvt]{Rodolfo G. Cionco\corref{cor1}}
\ead{gcionco@frsn.utn.edu.ar}
\address[rvt]{Grupo de Estudios Ambientales, Universidad Tecnol\'ogica Nacional, Col\'on 332, 
San Nicol\'as (2900), Bs.As., Argentina}
\cortext[cor1]{Corresponding author}
\author[focal]{Pablo Abuin}
\ead{pabloabu.g@gmail.com}
\address[focal]{Facultad de Ingenier\'ia, Universidad Nacional de Entre R\'ios, Ruta Prov. 11 km Oro Verde (3100), 
Paran\'a, Entre R\'ios, Argentina}

\title{On planetary torque signals and sub-decadal frequencies in the discharges of large rivers} 

\begin{abstract}

We explore the arguments presented in the past linking changes  in the 
angular momentum $modulus$, $|{\bf L}|$, of the Sun's barycentric orbit, with the 
discharges of Po River in Europe and Paran\'a River in South America, looking for any evidence regarding to 
a possible underlying physical mechanism. We clarify the planetary effect on solar torque presenting new analyses 
and results; 
we also improve prior results on Paran\'a River's 
 cycles finding significant spectral lines around 6.5 yr, 7.6 yr, 8.7 yr, and 10.4 yr. 
We show that the truly important dynamical parameter in this issue is the {\it vectorial} planetary torque.
Moreover, following the variations of ${\bf L}$ respect to the Sun's spin axis of rotation (i.e., a LS relationship),  
we found virtually the same Paran\'a River discharge peaks: 6.3 yr, 7.7 yr, 8.6 yr and  9.9 yr.
An analisys based on Magnitude Squared Coherence and Wavelet Coherence between Paran\'a River discharge and 
our LS relationship shows significant, although intermittently, coherence near 8-yr periodicities.  
Wavelet Coherence also shows big and significant regions of coherence inside 12-19 yr band. 
Our results ruled out classical tidal effects in this problem; but suggest that, if these rivers are trully 
related to solar barycentric motion, 
the physical origin of this connection might be related to a working solar spin-orbit interaction.

\end{abstract}


\end{frontmatter}



\section{Introduction}
\label{sec1}

The Sun's influence on terrestrial climate as a forcing mechanism 
 is not a new subject, and it has 
been studied for a long time 
\citep[e.g.,][and references therein]{H1910}. 
The solar internal equilibrium, implies a solar luminosity or total solar irradiance (TSI) that would be
nearly constant (often called ``the solar constant" in the past) or varying extremely slowly. 
However, the Sun poses a powerful dynamo which produces a complex and varying magnetic field, with surface 
manifestations such as sunspots, faculae, and magnetic network. This solar activity modifies the TSI reaching the Earth 
on different timescales 
through different mechanisms, the most obvious one observed is the 11-yr Schwabe cycle (cyclic variations 
in phase with the sunspot cycle and with amplitude of about $0.1\%$). Systematic variations on TSI are commonly 
the only solar forcing mechanism considered in tropospheric temperature and climatic evolution 
\citep[e.g.,][]{Grayetal2010,IPCC2007,Jonesetal2012}.
Nevertheless, the solar energy interaction with terrestrial atmosphere seems to be 
more complex than previously assumed, being the solar activity forcing's contribution to global and 
regional climate certainly underestimated \citep{Agnihotrietal2011,Soonetal2011,Stottetal2003}.

Changes in the Earth's energy budget due to incoming solar radiation, however,
 are not necessarily caused solely by Sun's own or intrinsic variations (i.e., are not 
necessarily originated by the own solar internal dynamics). For example, they can depend on factors external to 
the Sun, 
which in turn can modify the way in which the solar energy reaches the planet. 
The clearest example of these factors is the so-called Milankovitch forcing 
\citep[e.g.,][]{Cubaschetal2006} that is connected to the ever-changing Earth orbital parameters and other insolation 
quantites. 
The interaction of this forcing mechanism with the solar radiation is a purely geometric 
effect (i.e., it doesn't affect the intrinsic TSI emitted by the Sun).

There is also a growing body of evidence indicating that solar internal activity is influenced or modulated, at 
least at some extent, by planetary movements. This putative planetary influence in solar internal dynamics, that 
might be affecting the solar energy output, would now be physical.
The planetary hypothesis of solar cycle is an old idea that sporadically developed more than 100 yr ago, focusing mainly
on the more prominent solar
periodicity, namely the mean Schwabe sunspot cycle of 11.1 yr that is similar to the Jupiter orbital period of 
about 11.8 yr \citep{Brown1900,Wolf1859}. 
Since then, several works have described (mostly phenomenologically rather than physically), 
the possible planetary influence on solar activity, mainly through the modulation 
of sunspots cycles, as a proxy of the possible modulations on solar dynamo 
\citep{Charvatova2009,FairbridgeandShirley1987,Javaraiah2005,Jose1965,Landscheidt1999,WoodandWood1965}. 
Much more recently, this hypothesis
 has been revived with more specific evidence of the possible 
Sun-planets interaction \citep{Scafetta2012a,Scafetta2012b,ScafettaandWillson2013a,TanandCheng2013}, 
 highlighting several important aspects about the 
possible underlying physical mechanisms involved \citep{CC12,Scafetta2012a,WP10}. 
\cite{Abreuetal2012}, have suggested that the planets can be torquing 
the solar tachocline with periodicities similar to those observed in long-term solar activity proxy series, 
but \cite{CameronandSchussler2013} and \cite{PoulianovskyUsoskin}
 recently criticized their methodology and conclusions.

Several authors have also shown plausible dynamical planetary signals in climatic patterns on Earth, mainly in 
zonal-global temperature records and  auroral activity cycles 
 \citep{CharvatovaandStrestik2004,Landscheidt1987,Leal-SilvaandVelascoHerrera2012,Scafetta2014,Scafetta2012b,Scafetta2010,ScafettaandWillson2013b}. 
If the conclusions of these works are widely ``confirmed''
(and they were not exempt from criticism, see e.g., 
Holm, 2014a,b), a new perspective about the physical studies of solar action on climate should be considered. 
Recently, \cite{Shirley2014} have shown statistically significant evidence 
 relating changes in the orbital angular momentum of Mars (which is ruled by the other $N$-1 bodies of the Solar System) and martian 
 atmospheric circulatory anomalies, suggesting 
 the transfer of orbital momentum to rotational angular momentum.

In this sense, river flow dynamics, which can be considered a very good climate proxy, 
has also be linked to external, solar-planetary dynamics.
We are refering to the works by
\cite{AK11}; \cite{Landscheidt2000}; \cite{Tomasinoetal2000}; and \cite{Zanchettinetal2008}. 
The basic results of these works is that both Po and Paran\'a Rivers have sub-decadal periodicities that 
are related to similar 
periodicities in the variation of the inertial solar orbital angular momentum modulus ($L=|{\bf L}|$), 
i.e. periodicities 
in the so-called solar or planetary $T$ ``torque'' ($T$ = d$L$/d$t$). These studies have been limited to 
periodicities around  8 yr.

 These empirical results showed very good correlations between maxima and minima in annual series of $T$  
 and river
  discharge series ($D$ series) for Po and Paran\'a Rivers over the 20th century. 
 For Po River, \cite{Landscheidt2000} remarked that: 
``After 1933, all maxima of $|$d$L$/d$t$$|$ coincide relatively closely with outstanding discharge maxima, 
whereas all the $|$d$L$/d$t$$|$ 
minima mark discharge minima", and ``Before 1933, the relationship was reversed by a $\pi$ radians phase shift. 
It occurred when d$L$/d$t$ 
was exposed to a {\it perturbation} that deformed the sinusoidal course of the change in the Sun's orbital 
angular momentum" (the Italics are ours).
 We note that this result refers to the absolute value, $|T|$, not $T$. 
 In this approach, if a correlation exists between $D$ and $T$ series, only its extreme 
 values seem to be important, not the sign of $T$.
Another observed fact by
  Tomassino et al. (2000) is that the periodicity of the strongest spectral peak (8.7 yr) of the Po River 
  discharge is very close to the recorded mean length of the $|T|$ cycle reported by Lansdscheidt (2000). 
Both papers also reflected on the coincidence of river flows with the $|T|$ parameter and not $T$ itself.
On the other hand, the newer paper by \cite{Zanchettinetal2008} confirm a remarkable statistical correlation 
between $D$ series 
and extrema in $T$ series, in support of the earlier proposed planetary-Sun-climate forcing relationship. 
These authors also found a co-relationship between the 22-yr Hale sunspots cycles, Po's $D$ series spectra 
(with the main periodicity
 detected at 8.2 yr) and precipitation series ($P$ series). 
 Interestingly, Zanchettin et al. (2008) also mentioned the existence of ``perturbations" in $T$ series 
(a brief time when the sinusoidal amplitude of $T$ series drops significantly); 
 these perturbations  were also related by these authors to extrema in $P$ and $D$ series
\citep[][Fig. 4 on pg. 6]{Zanchettinetal2008}.

For the Southern hemisphere, \cite{AK11}, analysed solar signals and hydrological variability 
in Paran\'a River 
(the fifth most important river according to drainage area and the second largest drainage basin in South 
America)  roughly covering the last century. 
They, based on results from previous works, assume sub-decadal periodicities of Paran\'a River's $D$ series 
in the range 7-9 yr. Then using a multi-taper
 technique, they showed that Paran\'a's $D$ series and planetary $|T|$ series shared 
 significant 
 spectral power inside this particular band. This work is very interesting because of the huge drains area 
 addressed  (about $3\times10^6$ km$^2$)  which supposed
 to imply a strong climatic connection, and also because the phenomenological concordance between $D$ and 
 $|T|$ series. Indeed, from this spectral coincidence, the authors
  showed that Paran\'a's $D$ series and planetary $|T|$ series show a remarkable level of anticorrelation 
  \citep[][Fig. 4]{AK11}. 
 Notably, the ``perturbation" in $|T|$ series around 1935 seems also to be seen in Paran\'a's $D$ series. 
\cite{AK11} further shown that this 
  sub-decadal band is not present in sunspot time series, implying that solar irradiance would not be directly 
 related (at these timescales) with Paran\'a 
 River's discharge, suggesting that other physical mechanism linking solar activity and Earth atmosphere 
 should be sought after. 
 
 All these works related to Po and Paran\'a Rivers show empirical or phenomenological evidence in favour 
 of a possible direct solar-dynamics forcing on climate, and  an outstanding correlation between an 
 exclusive planetary origin parameter ($T$), and  rivers' $D$ series. 
Beyond the ``confirmation'' or disapproval of this possible relationship, these facts arises the question about
the possible causal link, i.e., the possible physical mechanisms present among planetary motions, 
the Sun's internal functioning and  Earth rivers dynamics. 

A first step in this direction is to go deeper into the previously published river's discharge and planetary 
torque relationships. 
At this point, several questions appear. The most satisfactory or convincing phenomenological relationship 
appears with $|T|$, 
but not directly with $T$. It is worth noting that \cite{Zanchettinetal2008} did use $T$ time series, but only 
refers to its maximum and ``perturbed" values when comparing with Po's $D$ series.
\cite{AK11} analysed  $|T|$ series and, as mentioned earlier, focusing only on the common 7-9 yr band between $|T|$ and 
$D$ series; they 
do not present detailed spectral peaks analysis in this band. First of all, examining Fig. 1 of \cite{AK11}, we note 
that the 7-9 yr band 
in $D$ series is significant at the 50-95$\%$ confidence level, 
whereas for $|T|$ series the signal is detected above the $95\%$ level. 
We reproduced these spectra (see Sec. 3 and \ref{sec4} for 
details on calculations) in Fig. 1 here. Note that at the significance levels of 50-95$\%$, both spectra share 
a rather broad spectral band, containing information from periods smaller than 7 yr and larger than 9 yr. 
Therefore,  we propose to study these relationships at sub-decadal timescale but without being limited to 
this 7-9 yr band, i.e., taking into consideration a broader spectral band.

The planetary torque acting on the Sun is, by definition, a vectorial quantity. 
Therefore, it is very important in this matter 
to assess the complete role of $T$ and ${\bf \Gamma}$, the bona-fide vectorial torque. 
In principle, there is no physical justification to take into account only the $T$ torque in the 
study of these solar-climate relationships. In addition it is absolutely unclear why the most important 
concordances 
occur with $|T|$ and not with $T$, furthermore, the absolute value $|T|$, has no immediate dynamic interpretation; 
 it is merely the absolute value of d$L$/d$t$, which is a scalar quantity.

An exploration related to planetary dynamics is imperative in order to known what are the planetary spectral 
frequencies involved at sub-decadal timescale and what is the
physical origin of the planetary signal against which we are comparing the rivers' flow variations. 

Classical tidal effects related to terrestrial planets
 (i.e. effects of deformations or departures in solar figure due to 
differential tide-generating forces), has been involved as possible underlying physical Sun-planets mechanisms, 
but in general, they were discredited \citep[e.g.,][]{OkalandAnderson1975}. Nevertheless, 
some planetary alignements involving terrestrial planets and also Jupiter, could be important respect to 
solar cycle 
\citep{Hung2007,Scafetta2012a}.  
Therefore, the involvement of terrestrial planets in this issue is interesting to discriminate.
 But, a set of other specific  questions  appeared at this point.
 For instance:  Which planets or planetary configurations are responsible of the spectral power observed in 
 the sub-decadal band? With respect to the 
abovementioned ``torque perturbations", why does it occur? What is its origin and interpretation in terms 
of planetary dynamics? Is it a merely descriptive definition or a clear dynamical effect?
 
In this exploratory study, we present original results that contribute to 
clarification of this interdisciplinary matter and to propose new hypotheses. The aim is to 
investigate the relevant planetary dynamics involved in the production of sub-decadal periodicities, using it 
as a starting point to look for any evidence regarding to a 
plausible physical Sun-planets mechanism related to these cycles. We also provide a discussion 
about possibles atmospheric drivers of this hypothetical Sun-rivers relationship. 

We pay particular attention on the Paran\'a's sub-decadal band because we have detailed data on the 
flow of this river, 
but because of the significant power spectrum shared with $|T|$ torque in addition to the alleged outstanding 
anticorrelation between $|T|$ and Paran\'a's $D$ series.
 We add that our results are also relevant for the  8.2-8.7 yr peak reported for Po River. 

In Sec. 2, we define the solar or planetary torque, specify its planetary components, and we show new 
features in $T$'s temporal evolution. In Sec. 3, we determine that the $|T|$'s power spectrum (at sub-decadal 
scale) is ruled by three strong spectral lines at about 6.6-yr, 7.9-yr and 9.9-yr periodicities due to Jupiter, Saturn 
and Neptune giants planets.
Sec. 4 is devoted to inquire about Paran\'a River's spectrum at sub-decal band. We found four statiscally 
significant lines around 6.5 yr, 7.6 yr, 8.7 yr, and 10.4 yr.
In Sec. 5 we show that the true important planetary dynamical parameter in this problem 
is ${\bf \Gamma}$, the vectorial torque, because these peaks in $|T|$ spectrum are due to the variations 
in ${\bf \Gamma}$'s modulus. Moreover, looking for the directional variations of ${\bf L}$ 
vector as seen from the Sun's spin axis of rotation, we were able to get a spin-orbit (LS) geometrical
relationship in Sec. 6. Analysing the oscillations of this LS relationship we found a set of frequencies  much  
closer to spectral features in Paran\'a and Po River flow analyses: 6.3 yr, 7.7 yr, 8.6 yr, and 9.9 yr. 
A Magnitude Squared Coherence (MSC) analysis (performed in Sec. 7) shows low, but significant 
(95\% confidence level) coherence between Paran\'a's $D$ series and our LS relationship  
near 8-yr periodicities. Wavelet Transform Coherence (WTC) confirms significant, although intermittently, 
coherence around 6-8.2 yr  band. WTC also detects very significant coherence between these signals 
inside the 12.3-19 yr band.
These findings are important because a solar spin-orbit coupling has been argued as a possible 
underlying physical mechanism linking Sun activity and planetary motions. We discuss this in the last section in 
addition to  possible atmospheric drivers of this hypothetical Sun-rivers relationship.

\section{The Sun's orbital angular momentum variation}

To clarify the nature of the planetary torque on the Sun, let us study the relationship between ${\bf L}$ 
and $T$: 

\beq
{\bf L} = M_{\odot} \, {\bf r} \times {\bf v},
\label{ec1}
\eeq

\noindent $M_{\odot}$ is the Sun  mass ($\simeq 2\times 10^{33}$ g), ${\bf r}$ and ${\bf v}$ are the position 
and velocity of the Sun in a barycentric reference system. Of 
course, ${\bf r}$ and ${\bf v}$ are due to the reflex barycentric motion produced by the other $N$-1 bodies of 
the Solar System:

\beq
{\bf r} = - \frac{1}{M_t} \sum_{j=1}^{N-1} \ m_j {\bf p_j},
\label{r}
\eeq

\beq
{\bf v} = - \frac{1}{M_t} \sum_{j=1}^{N-1} \ m_j {\bf \dot p_j},
\label{v}
\eeq

\noindent where ${M_t}$ is the total mass of the Solar System; $m_j$,  
 ${\bf p_j}$ and ${\bf \dot p_j}$ are mass, {\it heliocentric} position and velocity of the body $j$. 
 We have choosen a heliocentric system for positioning 
the bodies of the System because they are more commonly described on this reference system. Now, using this simple relationship: 

\beq
L^2 = {\bf L} \cdot {\bf L},
\label{ec2}
\eeq

\noindent we formally have:

\beq
\frac{ {\rm d}L}{{\rm d}t} = \frac{{\bf L}}{L} \, \cdot \frac{ {\rm d}{\bf L}}{{ \rm d}t}.
\label{ec3}
\eeq

\noindent The second factor in this multiplication is the planetary  vectorial  torque, ${\bf \Gamma}$,
 which now by virtue of Eqs. \ref{r} and 
\ref{v} has a clear dependence on planetary dynamics. Then, from Eq. \ref{ec3} we have:

\beq
T = \frac{1}{L}  \left( {\it \Gamma}_x L_x + {\it \Gamma}_y L_y + {\it \Gamma}_z L_z )\right .
\label{ec4}
\eeq

\noindent where ${\it \Gamma}_x$, ${\it \Gamma}_y$ and ${\it \Gamma}_z$  are the planetary vectorial torque components. In Fig. 2, 
the $T$ values are plotted from 1900-2013 at a monthly resolution. 
The solar barycentric dynamical parameters used in these calculations are obtained by using a code pack 
developed by 
the author, based on the planetary outputs of {\tt Mercury6.2} program 
\citep{Chambers1999} in high precision Bullirsch-Stoer mode. 
In Fig. 2, the contribution of giant and terrestrial planets to $T$ is clearly seen. 
The terrestrial planets has a non-negligible contribution because $T$ depends not only on solar position, but 
also on its velocity and acceleration. See \cite{WoodandWood1965} for a general comparison of solar dynamic quantities.

\subsection{The $T$ torque component of giant planets}

Fig. 3 shows $T$ torque component of giant planets, $T_G$. This figure reveals that $T_G$ long-term signal 
is basically ruled by the 
cyclical combination of Jupiter  and Saturn ($T_{JS}$) motions, i.e. by their synodic period (J-S), which for the
 last century
 we obtain a mean value of J-S = 19.84 yr. Of course, main extrema in $T$ (i.e., the peaks of $|T|$) appears at 
 semi-synodic period, i.e.,  the second harmonic of J-S frequency (i.e., each planetary 
quasi-alignments, conjunction-opposition), which is 9.92 yr in our simulations. 

Nevertheless, the effect 
of the other giant planets, Neptune and Uranus, is not negligible. 
For example, around 1990, $T_G$ shows a rapid 
variation, and this is due to an unusual, i.e., 
non-periodic extreme conjunction (very straight planetary alignment; see Javaraiah, 2005) of Jupiter with 
the other giant planets, which reduces drastically $T$ in a short timescale,  and produces
 a rapid but gradual angular momentum invertion of the solar 
barycentric orbit (SBO) and consequently, an extreme increase of its orbital inclination \citep{CC12}. 
As far we know, this is the first report on this ``anomaly" in $T$ evolution. 
But, we can also see the so-called ``perturbation" in $T$ series. Beginning in $\sim 1934$, 
Neptune counteracts notably the other three giant planets' torques. At this time, it ocurrs that Neptune is
approximately oposite to the other giant planets.
Other configurations generate a similar 
drop in 1970 ($\sim 36$ yr later, i.e., about one synodic period between Saturn and Neptune). 
Also the $T$ torque component combined from Jupiter, Saturn and Neptune 
($T_{JSN}$) is depicted in Fig. 3 (see Sec. 3 for additional explanation on this parameter).

At this juncture, our first conclusion is that
 the term ``perturbation" widely used with relation to solar-dynamics and river 
discharge relationship is purely descriptive: those drops in $T$ torque, obey the normal giant planet dynamics (it involves
 strongly Jupiter, Saturn and Neptune), they have about 36-yr
period and they are not produced by any anomalous situation or external force to the Sun-giant planets system.
 Indeed, as was mentioned earlier, a really unusual situation in $T$ occurs in shorter timescales 
  not related to terrestrial planets, that can only be 
clearly seen in a detailed $T$ torque representation using, approximately, monthly data.

\section{Sub-decadal band and giant planets periods}

Figs. 2 and 3 tell us that giant planets should produce the sub-decadal spectrum periodicities reported 
in $|T|$. Not all authors have used the same type of planets in their calculations, but on the other hand, 
it is important 
to assess if the short period terrestrial planets can contribute to some extent to 
 any shift or feature in the spectrum of $|T|$, taking into account the possibles above mentioned 
 tidal effects regarding to the solar cyle. This question,  can be easily assessed by means of spectral analysis. 
 We performed the same multi-taper
analysis as \cite{AK11}, using the same SSA-MTM-toolkit software \citep{Ghiletal2002} available at 
{\tt http://www.atmos\.ucla.edu/tcd/ssa/}, taking into 
account all the planets ($|T_{all}|$) and only giant planets ($|T_G|$). 
 For the analysis, we take one data point per year, as is obtained from our
simulations (no averaging was performed over the data). The planetary equations of motions were integrated 
using Bullirsch-Stoer scheme, from 1904 (to correspond with our Paran\'a River's data)
to 2012.  
We found (results not showed for brevity) that the effect of terrestrial planets in $|T|$ spectrum is negligible 
at this band, then all the signal at sub-decadal timescale observed in Fig. 1, is virtually produced by giant planets.

MTM method is a powerful tool for estimating low-amplitude harmonic oscillations in a relatively short time 
series with a high degree of statistical significance, but  could be a low resolution method 
 in terms of spectral lines determination. Therefore, we use maximum entropy method (MEM)
in order to assess more specific giant planets frequencies at this band.
MEM is a parametric, autoregressive (AR) method dependent on 
the order $M$ of the used poles, and must be 
lesser than certain maximum  in order to avoid spurious results, mainly at the extreme of Nyquist interval 
\citep[e.g.,][]{Penlandetal1991,Pressetal1992}. 
MEM has been very useful in order to isolate decadal frequencies in climatic series and also planetary 
frequencies (Scafetta 2012b-c, 2010). 
We used the same data set from 1904 to 2012 then $N = 108$ values were taken. For a rational estimate of pole 
numbers we begin with $M = 20$,
$M$=36 ($N$/3) and $M=54$ ($\sim N/2$). We assume, as customary in the literature, $N$/2 to be the maximum 
allowed value. 
The result is shown in Fig. 4, where the spectra of each pole number adopted are seen. 
There is no practical difference between $M$ = 36 and $M$= 54, but we decide to use
$M = 54$ because the peaks stabilize around this value, but also using a physics-based criteria: 
the 9.92-yr period due to semi-synodic period between Jupiter and Saturn is clearly resolved. 
We found, in addition to 9.92 yr signal, a significant period of 7.86 yr, and also a sharp peaks at 6.56 yr. 
Then, for a more stringent analysis, we repeat on $|T_G|$ the spectral calculation taking into account only giant 
planets dynamics starting from 800 A.D., assuming ergodicity and stationary planetary frequencies. 
The spectral peaks obtained are virtually the same confirming the results shown in Fig. 4 for the 
giant planets. Then, we performed a Lomb periodogram analysis using the latest, long-term JPL's planetary 
ephemeris (all the planets included) DE431 (Folkner et al, 2014; the Sun's acceleration was kindly provided by Dr. W. Folkner) 
from 13201 BC to AD 17191, and we found virtually the same peaks (6.64 yr, 7.78 yr, 9.92 yr): 
these ones, are the most important at sub-decadal scale. 

\subsection{Planetary origin of these peaks}
\label{sec3}

These cycles in $|T_G|$ are a robust result, then, 
we will attempt an assessment of their physical origin. As $T_G$ is the physical signal that responds to 
planetary dynamics; the half-periods of significant oscillations in $T_G$ should be the 
origin of the main periods in  $|T_G|$.
The 9.92-yr period is obviously the second harmonic of the synodic period of Jupiter 
and Saturn (J-S). On the other hand, Neptune, as we discussed before, and due to its long distance from the Sun, 
has a significant effect 
on $T$ signal. The measured synodic period with Jupiter (J-N) is 12.80 yr, therefore, its half value (6.40 yr) 
is strongly related to this 6.56 yr observed period, but it is perturbed by other planetary harmonics as   
S-N/3 $\sim 12$ yr and J-U= 13.81 yr.
 
With regard to the strong 7.86-yr peak, it is not obviously related to any evident planetary mean motion or synodic 
periods. Of course, we can obtain a lot of ``numerical'' relationships multiplying these observed periods by an 
arbitrary set of integer numbers, and many of these products will be near of giant planets synodic or 
semi-synodic periods and its harmonics. But also, and  more physics-based, 
 we can suggest an association with the mean observed period between pronounced extrema in $|T_G|$ series (as was 
 already mentioned in Sec. 1). For that, we calculate using our high resolution 15-day output series, 
 the average period between $|T_G|$ extrema and also the average time at which $|T_G|$=0 in differents ways, and 
 obtained differents values rangin from 7.72 yr to 8.28 yr. 
At a risk of being excesive, in the calculation 
of planetary periodicities we have taken two decimal places because we are comparing with a synthetic 
computational model, whose periodicities are known in advance. Then, to campare with river's spectral peaks, we use 
only one decimal place.  

The role of Neptune is determinant in the generation of the involved frequencies. 
As a check, we generate a new system, composed by 
Jupiter, Saturn and Neptune, with the same initial conditions as before. 
The $T$ torque coming from this subsystem ($T_{JSN}$) was already shown in Fig. 3. 
This figure shows that the $T_G$ is basically explained by these three planets, regardless of planet Uranus. 
Again, we search for the presence of periodicities and re-draw the spectral analysis over $|T_{JSN}|$. 
We obtained the following peaks through MEM calculations: 6.33 yr, 7.82 yr, 9.80 yr. This result shows
these three planets are indeed the main pysical cause of the above reported periodicities. 

These detected periodicities are the result of the mixing of these three planetary orbital frequencies in the 
formula that defines the $|T|$ parameter. As a last corroboration, we simulate a synthetic non-interacting 
system of three bodies in circular orbits with the periods of Jupiter, Saturn and Neptune, and calculate the 
corresponding $|T|$ formula. The resulting harmonic analysis of this $|T|$ parameter (not showed here), reveals virtually the same 
three periods, showing that they are produced only by the mixing of these three planetary periods.

Next, we are going to perform a comparison 
between $|T_G|$ frequencies and the 
rivers discharge frequencies at sub-decadal band. Regarding Po River, we remember that 
\cite{Tomasinoetal2000} and \cite{Zanchettinetal2008}, 
have reported a spectral line between 8.2-8.7 yr. 
Now, we want to assess the spectral peaks present in the Paran\'a River inside this broad sub-decadal band.

\section{Paran\'a River's peaks}
\label{sec4}

Significant spectral power between 8.4-9.2 yr was reported by \cite{RM98} and \cite{Krepperetal2008}
for Paran\'a River. 
We analysed Paran\'a River discharge series from Secretar\'ia de Recursos H\'idricos of Argentina 
({\tt http://www.hidricosargentina.gov.ar/acceso\_bd.php}), we follow \cite{AK11} by using
the same Corrientes (27$^{\circ}$28.5$'$S, 58$^{\circ}$50$'$W) 
gauging station data (see Fig. 5). To date (early 2014), there are daily data and 
monthly averaged data from January 1st, 1904 to August 31th, 2012. 
For the analysis we starts with monthly data and performed the analysis both using 
monthly and annual-mean data, the annual data were obtained averaging monthly data for the corresponding year. 
Therefore, we have taken into account $N = 1304$ (for monthly basis analysis) and $N = 108$ (annually averaged 
data). For the annual analysis, we have only taken into account data till 2011 because 2012 datasets are incomplete.

Results from the annual and monthly data sets are comparables. Fig. 6. shows the results for 
the monthly data. The MEM spectra show prominent peaks at  6.5 yr, 7.6 yr, 8.7 yr and 10.4 yr. 
It seems that Paran\'a River has more peaks in this spectral zone than $|T|$ torque of giant planets. Strictly 
inside  the 7-9 yr band only one peak around 7.6 yr is commonly shared between planetary $|T|$ and the river 
record. 

Now, we address the problem of 
the significance of these peaks, we want to be sure that these peaks are statistically meaningful signals, i.e., 
that we are not fitting a subtantial amount of 
noise in the AR model. The MEM spectra can be constrained with red-noise spectra using a Monte Carlo
 permutation test, as was made by \cite{PIRT05}. The {\tt MAXEMPER} software coming from this publication 
({\tt http://www.iamg.org/ind\ ex.php/publisher/articleview/frmArticleID/118})
 evaluates 
the statistical significance of the spectral estimates using the mean power spectra of the $N'$-th random 
permutation. This mean spectra 
is using for testing the null hypothesis from which the spectra of the random permutations are sampled. 
The outputs include the achieved 
significance levels of the power spectrum estimated for each frequency. 
Fig. 7 shows the confidence level of the signal that reaches 
$95\%$ and higher. 
Panel a) shows significant power in $\sim$ 7-10.5 yr band for $M = N/4$, panel b) depicts the same 
for $M = N/3$ and 
the last one shows the significant spectra for $M = N/2$. 
We see how the central 7-9 yr band is refined in more detailed spectral lines: 
in addition, the 6.5 yr line is increasing in importance reaching the $99\%$ confidence level, 
the 10.4 yr peak is only visible with 
a (reasonable) value of $M = N/2$, but significant at $95\%$ level against red-noise spectrum.

In order to carry out a even more stringent analysis and eliminates red noise of monthly series, 
we performed a Singular Spectrum 
Analysis (SSA) by using the same SSA-MTM-toolkit. 
We used a conservative $M = 130$ spectral window 
(enough to resolve 6-yr periodicities) with 11 temporal empirical orthogonal functions (T-EOFs); then, 
we selected those T-EOFs which are 
oscillatory according to the corresponding test (specificaly, strong FFT). 
Hence we retain three oscillatory pairs (T-EOFs 1-2, 4-5, 6-7), 
and performed the reconstruction using the corresponding principal components. Then we used 
this reconstructed-filtered signal 
to perform MEM. The result in Fig. 8 yields basically the same spectral peaks as shown in Fig. 6. 
Therefore we accept that these detected peaks around 6.5 yr, 7.6 yr, 8.7 yr and  10.4 yr, with $M=N/2$, 
are statistically significant. 

This refinement of the MEM spectrum also shows important power above 10-yr periodicities, i.e., around 12-13 yr 
and 18-19 yr. {\tt MAXEMPER} permutation test shows that this periods are significant (respectively) at 95$\%$ and 
$99\%$  confidence level. In Sec. 6, WTC coherence analysis will show that this last spectral zone might play an important 
role in this problem.   

\section{The scalar torque modulus, $|T|$, and vectorial torque, ${\bf \Gamma}$, variation}
The importance of the vectorial torque ${\bf \Gamma}$ should be addressed 
and highlighted. So far, torque parameter, $T$, is the 
only physical quantity cited in this issue, but $T$ is a scalar that only take into account $L$ variations. 
 ${\bf \Gamma}$ = d${\bf L}$/d$t$ is 
the quantity that measures the total angular momentum variation of the inertial movement of the Sun. 
Then, let us to analyse the torque strength or vectorial torque modulus $|{\bf \Gamma}| = {\it \Gamma}$. 
Following Eq. \ref{ec3}, we have: 

\beq
|T| =  {\it \Gamma} \,  |{\rm cos} ({\bf \Gamma},{\bf L})| 
\label{ec6}
\eeq

\noindent the argument of the cosine function is the angle between ${\bf \Gamma}$ and ${\bf L}$ 
associated directions. In what follows, accordingly with our findings, we discard the contribution of the 
terrestrial planets and only consider the influence of the giant planets (i.e, $T_G$).

As the planetary movement is quasi-planar, and the solar {\bf L} vector is basically directed 
to $z$-axis of the inertial system 
for long time intervals (see Cionco and Compagnucci, 2012, for a detailed evolution of SBO inclination), 
we can expect that 
 that cycles in ${\it\Gamma}$ due to giant planets, to be the real origin of these sub-decadal frequencies in $|T_G|$. 
For testing this idea, we calculate the MTM and MEM spectrum of ${\it \Gamma}$.
We found that both MTM spectra and the MEM peaks (result not showed for brevity), are coincidental and 
agrees with $|T_G|$ spectra previously obtained in Sec. 3. 
An analysis of cosine function of Eq. \ref{ec6}, do not show any significant power in the sub-decadal band.
Also, a periodogram analysis performed over ${\it\Gamma}$ derived from JPL's DE431 long-term ephemeris, yields 
the same peaks reported in Sec. 3.
The reason that previously analyses only take $|T|$ into account but do not analyse the vectorial torque in this 
problem is unknown to us, at least on physical grounds. 
Analysing the torque modulus the same relationships should have been found and this seems to be more clear 
and natural. 

After confirming that cycles in ${\it \Gamma}$ produces the observed periods in 
$|T|$ torque, we now return our attention to 
the directional variations of ${\bf L}$ in space. In addition to that oscillations in ${\bf L}$ modulus, this 
 vector has a precessional-like and nutational-like movements around $z$-axis of the inertial system. 
We note that the inclination of the SBO has customary variations of few degress 
($\sim$ 1-6 deg) (Cionco and Compagnucci, 2012).
This kind of ``nutation in obliquity'', is certainly of very low amplitude, 
but the precession-like 
movement of the SBO orbit is easily seen following the evolution of the ascending node ($\Omega$) of the SBO, 
this is the same 
precessional movement of ${\bf L}$ around $z$-axis of the inertial system (because ${\bf L}$ is normal to 
the orbital plane by definition, and the nodal line is perpendicular to $z$-axis of the inertial system). 
Fig. 9 shows $\Omega$ variations in the studied period. 
We can clearly see as $\Omega$ varies between 0 and 360 deg and 
also performs bounded oscillations, alternatively, through the time. 
Nevertheles, for simplicity sake, we will refer 
to this precession-regression movement as a precesional change of ${\bf L}$ around inertial $z$-axis.

At this moment, it is important to note that planetary tidal effects and spin-orbit coupling have been the main 
lines of inquiry about the 
underlying physical mechanism in the planetary hypothesis of solar cycle. The giant planets frequencies involved 
in the sub-decadal band ruled out classical
 tidal effects in this problem, i.e., tidal effects involving terrestrial planets.
  As we discussed, the time evolution of ${\bf L}$ 
 is not trivial. The angular momentum vector has basically its most important component along $z$-axis of the 
 inertial system, it has continuous oscillations, precessional-like changes and brief 
 and sporadic rotations of about one year, when 
 solar orbit is gradually inverted (Cionco and Compagnucci, 2012). 
This means that planetary torque also produces appreciable
 changes in the direction of angular momentum in the inertial space and also more specifically, with 
 respect to the Sun's spin axis (${\bf S}$). 
 Authors such as \cite{Chang09}, \cite{Java15}, Javaraiah (2005) and \cite{Juckett2003} have analysed Sun's rotation and 
 sunspots distribution 
 showing similar periodicities found in $T$ series (i.e., $\sim$ 8-9 yr). That, has been considered as the evidence of a 
 spin-orbit 
 coupling in the Sun. \cite{PMAC11} have taken this mechanisms into consideration and have argued that 
 exoplanetary systems could be a useful environment for further
 testing and corroboration of a solar spin-orbit coupling hypothesis. These authors have only taken into account 
 variations in spin rate 
 of the Sun's rotation axis respect to variations in $T$ torque. 
We  have already calculated the frequencies involved in the variation of the modulus of ${\bf \Gamma}$ which, in 
turn, rules $L$ variations. Now, we are going  
to evaluate the directional change of ${\bf L}$, but with respect to the Sun's spin axis, ${\bf S}$.

\section{ ${\bf L}$ and ${\bf S}$ geometrical relationship}

The simplest way to relate  ${\bf S}$ and ${\bf L}$ is by studying
 the orientation of ${\bf L}$ respect to ${\bf S}$ through the time. 
 As far as the author knowledge, this geometrical relationship has not been performed or studied in the literature. 
 \cite{Juckett2000} studied the normalized 
projection of ${\bf r}$ towards ${\bf S}$, he proposed this geometrical projection (which is basically the $\cos({\bf r},{\bf S}$)) 
as an ``spin-orbit indicator'' 
and searched in this indicator for giant planets signal at over-decadal and longer timescales. 
Nevertheless, the use of ${\bf r}$, instead of ${\bf L}$, could be inadequated. Although ${\bf r}$ is 
perpendicular to ${\bf L}$, by 
definition (then variations in ${\bf r}$ can express variations in ${\bf L}$), ${\bf r}$ evolves with Sun's 
orbital motion, independent of ${\bf L}$ 
position in the inertial space, then, that proposed indicator, has other frequencies not related to 
the relative evolution 
between ${\bf L}$ and ${\bf S}$.
Moreover, taking into account that ${\bf S}$ is neither fix in the inertial 
system, the situation is even more complex in reality because it evolves secularly.

To accomplish this LS relationship task, the ecliptical-inertial system will
be the linkage between the angular momentum and the Sun associated coordinate system.
We adopted the Inertial Heliographic System (IHS)
attached (but not fixed) to the Sun \citep{Burlaga1984,FranzandHarper2002}. 
In this system the $z$-axis is defined along the Sun's spin axis 
and the $x$-$y$ plane coincides with the solar equator. Then, the system is defined with 
respect to the inertial system by means of the Sun's obliquity $\epsilon_S$,
and the longitude $\Psi_S$ of the intersection between the ecliptic and the solar equator (Fig. 10).
 We adopted their values referenced to the epoch  J2000.0 \citep{FranzandHarper2002}:

\beq
\epsilon_S = 7^{\circ}.25
\eeq

\beq
\Psi_S = 75^{\circ}.76 + 1^{\circ}.397 \, (t_0-t)
\eeq

\noindent where $t_0-t$ is the fraction of Julian century from J2000.0. Then the $x$-axis of the IHS 
system (${\bf Xs}$) is 
the intersection of 
the solar equator and the ecliptic of the corresponding epoch. Therefore, we linked the inertial and the 
IHS system by means of the following rotation matrix product:

\beq
(x_s,y_s,z_s)^{t} =  {\bf R}(\epsilon_S)  {\bf R}(\Psi_S) (x,y,z)^{t}
\eeq

\noindent $x_s$, $y_s,$ and $z_s$ are the components in the IHS system of the $(x,y,z)$ vector in 
the inertial system. 

The positioning of ${\bf L}$ with respect to ${\bf S}$ can be acomplished as usual in spherical astronomy, i.e., 
using two spherical angles 
associated with two orthogonal directions on the  
celestial sphere: $\delta$ measured from the Sun's spin axis toward ${\bf L}$ (a kind of colatitude angle), and 
$\alpha$, measured from ${\bf X_S}$ towards the arc of  
great circle that connect ${\bf S}$ and ${\bf L}$ (a longitudinal angle; see Fig. 10). Both angles are defined by 
the following expressions:

\beq
{\rm cos}(\delta) = \frac{Lz_s}{L}
\label{cosd}
\eeq

\beq
{\rm cos}(\alpha) = Lx_s \,  [L^2 -  Lz^2_s]^{-1/2}
\label{cosa}
\eeq

\noindent where $Lz_s$ and $Lx_s$ are the $z$ and $x$-component of ${\bf L}$ in IHS. Then, we follow the 
evolution of ${\bf L}$ respect to ${\bf S}$ for 
the same period consistently with Paran\'a River data. We analysed from 1904-2012 A.D., and then performed 
spectral analysis to 
both positional angles. Fig. 11 show the evolution of $\alpha$ and $\delta$. The co-latitudinal angle $\delta$ 
shows very small amplitude 
oscillations but a sudden increase around 1990 because of the above-mentioned orbital invertion (${\bf L}$ 
invertion). The longitudinal 
angle $\alpha$ shows more important oscillations of about 20 deg in amplitude, with marked peaks each 
$\sim 38$ yr, 
(which are also visible, but lesser pronounced, in $\delta$). These secondary peaks occur at 
 Jupiter, Saturn and Neptune alignments (remember that the  recorded mean S-N period is $\sim$ 36 yr). 
The exceptional four giant planets alignment of 1990 is also seen. 

Consequently, MEM and MTM analyses of $\delta$ do not show any significant spectrum (results not shown here), 
but $\alpha$ evolution shows very significant spectral power in the sub-decadal band. 
This spectrum is showed in Fig. 12, where Paran\'a MTM spectrum is also depicted for comparison. 
In this figure, we have plotted two vertical lines delimiting the spectral zone in wich the power of $D$ is larger than $50\%$ red 
noise significance level. Interestingly, we note 
that the Paran\'a spectrum is much more similar to $\alpha$ spectrum than the $|T|$ (or $|T_G|$)  spectrum at sub-decadal 
band (compare with Fig. 1).
Indeed, MEM spectrum of $\alpha$ 
shows significant peaks, coincidents with Paran\'a River periodicities (Fig. 13). We can see four significant peaks at 
6.3 yr, 7.7 yr, 8.6 yr and 9.9 yr. Respect to $|T_G|$ spectra, these periodicities are certainly the same as Fig. 4, 
with the addition of a peak in 8.6 yr. Using the same methodology applied in Sec. 3 to identify the peaks 
in $|T_G|$ signal, we conclude that these peaks are due only to Jupiter, Saturn and Neptune, and this peak at 
8.6 yr also comes from the mixing of these planets' periods in the formula (12).

Two peaks of 7.7 yr and 8.6 yr practically coincides with Paran\'a peaks inside 7-9 band (7.6 yr 
and 8.7 yr). Particularly, this 8.6-yr period is very close to the Po and Paran\'a Rivers strongest peak. 
The percentage difference 
between Paran\'a River's spectral peaks and $\alpha$ periodicities in the sub-decadal band is lesser than $5\%$. 
At this point, it is interesting to note that, taking into account the Nyquist theorem, the approximate 
error in the determination of periods ($p$) in Paran\'a River's series ($L=108$ yr) is $\sim p^2/(2L)$ 
(Tan and Cheng, 2013); then, the 10.4-yr peak is ranging from 9.9 yr to 10.9 yr; i.e, $\sim 5\%$ of error.   
Then, in the context of the previous related (empirical) studies, we 
have found a stronger phenomenology; and this, can be linked to a physical hypothesis 
of Sun-planets interaction, i.e., a putative solar spin-orbit coupling effect depending on this LS geometry.
Now, we are going to evaluate if these $\alpha$ and $D$ signals have any degree of spectral coherence 
as to take into consideration our proposed LS relationship in a plausible scenario of 
Sun-rivers physical relationship.

\section{$\alpha$ and $D$ spectral coherence analysis}

Fortunately, we have several methods to evaluate the frecuency-domain similarity or spectral coherence between two signals. 
The coherence function or MSC can be evaluated  by several methods 
\citep[see e.g.][]{Benesty,Holm2014a,Scafetta2014}. 
After several initial trials with synthetic signals, we have decided to use the Minimum Variance Distortionless 
Response (MVDR) 
\citep{Benesty} for evaluating MSC because it is more accurate (or restrictive) in the isolation of 
the significant frequencies than the usual routines based on Welch's periodogram method. 
We used the Matlab MVDR implementation of 
\cite{Benesty} 
({\tt http://www.mathworks.com /matlabcen\ tral/fileexchange/9781-coherence-function/content/cohe\ rence\_MVDR}).
MVDR is also easier to use: it only depends on the number of filters ($K$) and their window lenght ($L$).
To get better resolution than annually data, we have analysed both $\alpha$ and $D$ data on monthly basis 
($N = 1304 $) and 
used $L=108$ yr  \citep[see e.g.][]{Benesty,Scafetta2014}. Tha planetary data was simulated taking 12 values per year of 365.25 yr; no 
averaging was performed on it. The data sets were detrended and $z$-standarized.

For evaluating the MSC coherence between $\alpha$ and $D$ signals and its confidence levels, we found the MVDR coherence of the original signals and 
also performed a Monte Carlo permutation test in the strategy of \cite{PIRT05}, i.e., finding the Achieved Significance Level for each 
sampled frequency. It is the probability to get, by chance, a surrogate MSC signal grater than the original MSC signal. 
We performed 5000 permutations with no restrictions, so a stringent withe noise hypothesis have been used. 
Finally, it is straightforward to obtain de Achieved Confidence Level (ACL) for each frequency-periodicity.  
Fig 14 shows this: the solid line is the ACL of the MSC calculation obtained in the Monte Carlo test.   
The dashed line is the 95\% confidence level. 
Therefore, MSC between $\alpha$ and $D$ signals, are statistically significant at 95\% confidence 
level around 8-yr periodicity. This is an encouraging result, despite its marginal significance. 
This could indicate intermittent coherence along the time. 

To study this, we used an independent method based on Wavelet Transform \citep[squared WTC, ][]{Grinsted}.
We used the corresponding software at 
{\tt http://noc.ac.uk/using-science/crosswavelet-wavelet-coher\ ence}. The result is showed in Fig. 15. 
 WTC detects high level of coherence between 6-8.2 yr. This coherence 
seems to be stable until 1930, then later return about 1990, but specifically around 8-yr periodicity. 
These signals seems to be out of phase ($\sim -45$ deg) with some tendency to be in-phase. 
Another more important island of coherence appears inside 12-19 yr band (specifically between 12.3-19 yr), 
where the signals are out-of-phase/anti-phase.
This spectral zone is very signifcant for $\alpha$ signal, it is related to 
Jupiter orbital period, J-S, J-N, S-N/2, etc.    
Spectral power around 13 yr and 18 yr was also observed in the SSA filtered MEM analysis (Fig. 8), and confirmed by 
Monte Carlo permutation test. Therefore, squared WTC appears also to detect very significant coherence at this 
over-decadal spectral band.

\section{Discussion and concluding remarks}

In this effort, we have re-evaluated key dynamical aspects related to the evidence presented in the past 
linking solar inertial motion and discharges from Po and Paran\'a Rivers. 
We have explained the cycles and the physical origin of the signals present in the oscillations of $|T|$, the 
most important parameter taken into account in these empirical evidences. We showed the existence of three 
clear frequencies, originated in the variations of the vectorial torque modulus (${\it \Gamma}$) ruled by the 
planets Jupiter, Saturn and Neptune. We also improved prior results on 
Paran\'a River discharge periodicities, showing the existence of four significan spectral peaks.

Since the detected importance of vectorial torque in this problem, 
 we found basically the same Paran\'a 
discharge spectral peaks in the spectrum of the longitudinal variations ($\alpha$ angle variations) 
of ${\bf L}$ with respect to ${\bf S}$ (which are ruled by the same set of planets).
Having ruled out the tidal hypothesis in this problem, this result stresses the importance of the 
vectorial torque, taking into account the hypothesis of a possible solar spin-orbit coupling effect.
In addition, Po River also shows a sub-decadal spectral peak (8.2-8.7 yr) more similar to $\alpha$ spectrum peaks than $|T|$ spectrum peaks. 

Beyond these spectral matches and phenomenologycal coincidences, we were able to obtain significant mathematical 
coherence between both $\alpha$ and $D$ signals, which is a very encouraging result. 
These findings suggest that, if these rivers are trully linked to solar dynamics, 
the physical origin of this connection might be related to a working solar spin-orbit interaction.
Nevertheless, to describe the complete scenario, we must seek for a terrestrial linkage between Sun-planets 
interactions and river dynamics. This ``atmospheric driver'' would be the directly responsible for perturbing
 river dynamics, whereas 
it respond to external, solar fluctuations. This putative driver, should show a very good coherence with 
both Sun dynamics and river dynamics at particular spectral bands; but certainly, it would act as a filter, 
enhancing particular cycles and attenuating others, following its own dynamics, which 
can also produce different responses 
in different rivers, taking into account their locations on the Earth.

The North Atlantic Oscillation (NAO), was proposed as a driver of this possible Sun-rivers connection.
Zanchettin et al. (2008) have showed that NAO is correlated with both  
solar activity and Po River discharge at sub-decadal time-scale. Scafetta (2010) suggests a NAO and solar inertial 
motion relationship at multi-decadal timescale. 
Oscillatory modes between 7-8 yr have been detected by \cite{Palus&Novotna} 
 in  NAO and geomagnetic index. Also, \cite{Georgievaetal12} have shown strong connection among heliospheric 
 activity, geomagnetism and NAO oscilations.
 Therefore, changes in solar magnetic activity, NAO oscillations and variations of 
 hydrological  patterns related to  Po River, are spected to be strongly connected.
 
 By other hand, Robertson and Mechoso (1998) have related at sub-decadal time-scale, anomalous cool events 
in Tropical North Atlantic (TNA) with high South-America rivers runoff. This seems to have motivated
 \cite{AK11} to 
propose a possible relationship NAO-Paran\'a. 
A more difficult issue to address is the influence that NAO (as a whole atmospheric-oscillatory
system) might have over South-America, at this sub-decadal time-scales. The TNA affects 
intertropical South-America via, e.g., decadal variability of the summer monsoon system (Robertson and Mechoso, 
1998); then, it is possible that TNA acts as agency between South-American precipitation regime and NAO.

Paran\'a's basin presents very inhomogeneous 
regions with different hydrological patterns. The average annual precipitation decreases from east to west 
(i.e., as we move away from the Atlantic Ocean), but also from north to south \citep{dpp}.
\cite{Pintoetal13} have recently showed that thunderstorm days in Brazil are correlated
to solar activity; but a very specially feature arise from their work: the most southern data set used, coming 
from Porto Alegre city station, shows a broad spectral peak around 8 yr. 
 They related their findings to possible magnetic activity changes in Earth atmosphere. 
Porto Alegre city is almost at the same latitude than Corrientes gauging station (Fig. 5). 
Therefore, it is imperative to study the rainfall regime in other cities of the Paran\'a basin, looking for 
these sub-decadal periodicities, and particularilly, a possible north-south gradient with these periodicities  
\citep[taking into account][results]{Pintoetal13}. 

 South-east of Brasil is the centre of the South
Atlantic Magnetic Anomaly 
(SAMA). It produces the sinking of charged particles trapped in atmospheric belts \citep{Pintoetal91}. 
This anomaly (unique phenomenon in the world) affects almost all Paran\'a basin, but is stronger 
at the south-east of the basin. Therefore, solar, geomagnetic 
and atmospheric activity should be carefully investigated, as a whole, in this region. 
We think that Sun-climate relationship with magnetic activity variations, should be considered as 
the potential connection of these possible Sun-river relationships. Sub-decadal variations of charged 
particles coming from the Sun (for instance, 
driven by a solar spin-orbit interaction), could interact with Earth atmosphere and geomagnetism, producing   
complex climatic patterns.

Earth climate variations with relation to a possible solar spin-orbit coupling have also been argued 
by \cite{Shirley2009}, taking into account solar meridional fluxes changing velocities. 
Geomagnetic activity is strongly dependent on solar dynamo, through changes or alternations in poloidal and 
toroidal fields \citep[e.g.,][]{Georgievaetal12}, then a possible relationship between heliospheric parameters 
and the Earth atmosphere, and this as a trigger of hydrometeorological signals, is expected.
A spin-orbit coupling hypothesis was called upon to explain some phenomenological concordances  between $L$ 
variations and solar activity, through solar rotation and functioning of a magnetohydrodynamical dynamo. 
 The original idea seems to be first proposed by Jane \cite{Blizard1981}, and has since been taken into
 consideration also by 
\cite{Javaraiah2005}, \cite{Juckett2003,Juckett2000}, \cite{PMAC11}, \cite{Shirley2006}, \cite{Shirley2014},
 \cite{Zava}. 
The idea is that planets can transfer orbital momentum to the Sun's rotational angular momentum, and 
this variation could interact with the solar dynamo through a putative mechanism. Therefore, some 
part of the orbital angular momentum could be transferred to spin angular momentum. The inverse can also 
be true as was shown by Javaraiah (2005). As was mentioned Javaraiah (2005) and Juckett (2003) show evidence that
certain periods in Sun's rotation are very similar to periods in $T$ power spectrum, especially 
the 8-yr periodicity. Here, we have presented a complementary side of this possible physical phenomena, i.e., 
the study of the {\it directional} variations between ${\bf L}$ and ${\bf S}$.
It is also interesting to note that, oscillations of $\sim 8.5$ yr have also been measured in solar 
cycle \citep{Rozelot1994}.
Periods from 6-8 yr has been found in drifts of latitudinal bands of near-equal rotational velocity in the 
Sun \citep{Makarovetal}.

At present, there is still no clear physical mechanism to explain how this transference of angular momentum 
might be achieved. 
In planetary dynamics, spin-orbit coupling refers to spin-orbit resonance 
\citep[see e.g.][Chap. 5 for this subject]{MurrayandDermott1999}, i.e., a commensurability 
that appear between the spin rate of a body and its orbital period. In general, the spin axis of the body 
is considered parallel to its orbital angular momentum vector (spin perpendicular to the orbit or zero 
obliquity approximation).
 The mechanism behind this coupling is the tidal friction
originated by a planet over a satellite \citep{GoldreichandSoter1966}, or by the Sun to the planets
\citep{GP66,PealeandGold1965}. Goldreich and Soter (1966), argue that tides
raised on the Sun by planets have virtually no effect on the rotation and orbit of the Sun.
 But this does not preclude the fact that certain internal solar dynamics can be susceptible 
to external gravitational modulation and perturbation (Scafetta, 2012a,b; Wolff and Patrone, 2010). 
Our problem at hand is more complex, because the solar orbital angular momentum has great variations (in 
modulus and direction), unlike of the 
orbital angular momentum variations of the planets. A model of solar spin-orbit coupling was out of the scope 
of this work; we only arrived at a spin-orbit relationship following the dynamics involved in this problem. 
But we can say that any model of solar spin-orbit
 interaction should consider 
\citep[see e.g.][]{Peale2005}: a) an adequate expansion 
of the gravitational potential of the Sun that accounts for the Sun's permanent figure, and the calculation of 
the corresponding planetary torque; b) the tidal torque of the planets and; c)
the frictional torque coming from 
different solar internal zones (tachocline, convective envelope, etc). These zones 
have different shapes/forms (e.g., ellipticities) and then, they could precesses at different rates \citep{Poincare}. 
Therefore, these 
ingredients are essential to a detailed description of the Sun's spin axis evolution (tilt and rate) and its possible coupling 
with the orbital angular momentum at different time-scales. Our findings also suggest that, at least, 
we must consider the contributions of planets Jupiter, Saturn and Neptune. For example,  \cite{Changetal2012}, 
recently shown how a young star can modify its axial tilt by the magnetic torque that arises from the
Ohmic dissipation in a Hot-Jupiter planet system, at the expense of the spin-orbit energy (see Eq. 17 of that
 paper for a spin-orbit coupling expression). 
The idea of Abreu et al. (2012) about the planetary tidal
interaction in the tachocline, which has been modelled as an ellipsoidal figure, 
is a good candidate for spin-orbit coupling interaction, and should further be studied in this context. 

Of course, a lot of work is needed to study several aspects of these processes and hypotheses; specially, to 
analyse another rivers and the suspected atmospheric drivers. We are working on this, and the results 
will be the subject of another paper.

\section*{Acknowledgements}
To be provided


\newpage

\noindent {\bf FIGURE CAPTIONS} (ALL FIGURES IN BLACK-WHITE)\\

Fig. 1. MTM spectra comparison of Paran\'a $D$ series and  planetary $|T|$ torque. The vertical doted lines 
mark the common 7-9 yr band considered 
by \cite{AK11}. Inside this band, the spectral power of $D$ is significant at 50-95$\%$. The simple inspection 
of this figure shows that $D$ signal is significant at this levels in a broader sub-decadal band.\\

Fig. 2. Planetary $T$ torque from 1900 to 2013 A.D. Dashed line: all the planets included; solid line: only giant planets. 
Physical units: solar mass (Ms), astronomical unity (AU) and days (d).\\

Fig. 3. Comparison between $T$ torque of only giant planets ($T_G$), $T$ from Jupiter and Saturn ($T_{JS}$), and from Jupiter, Saturn and Neptune ($T_{JSN}$) 
subsystem planets. 
Surpraisingly, the rapid variation (lesser than 1 yr) in $T$ signal around 1990 is due only 
to giant planets. The quasi-conjunction of Jupiter with other Giants forces to $T$ to be near zero. 
 The $T$ ``perturbation''around 1935 is due to the normal dynamics that strongly involve Neptune planet.\\

Fig. 4. Maximum entropy (MEM) spectra for $|T_G|$ with differents number of  poles $(M$) used.\\

Fig. 5. Paran\'a's basin. The main rivers of Paran\'a system are showed. Corrientes city in Argentina is marked. 
Other South-American cities in Uruguay, Brazil, Bolivia and Paraguay are also indicated.\\

Fig. 6. MTM and MEM Spectra of Paran\'a's monthly $D$ series (1304 data values used).\\

Fig. 7. Results of the Monte Carlo permutation test ({\tt MAXEMPER} software) performed to Paran\'a monthly 
$D$ series, showing the significance 
of these peaks against red noise null hypothesis. The pole order $M$ is showed (number of data $N$= 1304).\\

Fig. 8. MEM spectrum of the filtered $D$ monthly series showing the disappearance of noisy-peaks ($M = N/2$). 
The small window shows in detail that the same four peaks appear in the sub-decadal band. For comparison, the arrows indicate 
the peaks of the raw-annual data.\\

Fig. 9. Ascending node evolution of the solar barycentric orbit (one point per year), the same oscillations are performed
 by $\bf{L}$ vector around the $z$-axis of the inertial system, in this case, the ecliptical J2000.0 system.\\

Fig. 10. Sketch of the IHS system showing the solar equator ($Qsolar$), the ecliptical plane (the $x$-$y$ plane of the inertial system) 
and its pole ${\bf Z}$ (the $z$-axis of the inertial system). The angles of orientation $\epsilon_s$ and $\Psi_s$ 
between the IHS and the inertial system are indicated (the orientation between ${\bf S}$ and ${\bf Z}$ was largely exagerated by convenience). 
The ${\bf L}$ vector has a precessional-like movement around $z$-inertial axis (it was idealized by the shadowed ellipse). 
The position angles of ${\bf L}$ respect to ${\bf S}$ ($\alpha$ and $\delta$) are indicated with double arcs.\\

Fig. 11. $\alpha$ and $\delta$ evolution in the studied period. The most important peaks occur each $\sim$ 38 yr, due to Jupiter, Saturn 
and Neptune alignments. The greatest peak around 1990 involve the four giant planets.\\

Fig. 12. MTM spectra of Paran\'a and $\alpha$ series. 
The vertical lines mark the band at which the spectral power of $D$ is significant at 50-95$\%$. 
Surprisingly, in this sub-decadal band four peaks seem to be very similar in both series. 
Certainly, Paran\'a spectrum  is more similar to $\alpha$ spectrum than $|T_G|$ spectrum. 
Also, a prominent bi-decadal band is also seen in $\alpha$ spectrum. \\

Fig. 13. MEM spectra of Paran\'a $D$ series and $\alpha$ series. It confirms the
 existence of four peaks. Therefore in the sub-decadal band Paran\'a's $D$ series has virtually 
 the same frequencies than $\alpha$ series (the porcentual error is lesser than 5$\%$).\\

Fig. 14. Achieved Confidence Level (ACL) of the MSC (MVDR method), with Monte Carlo permutation test,
 between Paran\'a $D$ series and $\alpha$ series. Periodicities 
around 8 yr are statistically significant at 95$\%$ confidence level. \\   

Fig. 15. Squared WTC coherence between $D$ and $\alpha$ series. It confirms the coherence's intermittency around 
8-yr band. The horizontal dashed lines are the approximated bounds of 95$\%$ confidence level zones, they   
marks periodicities between 6-8.2 yr and 12.3-19 yr.

\newpage

\begin{figure}
\includegraphics[scale=1.1]{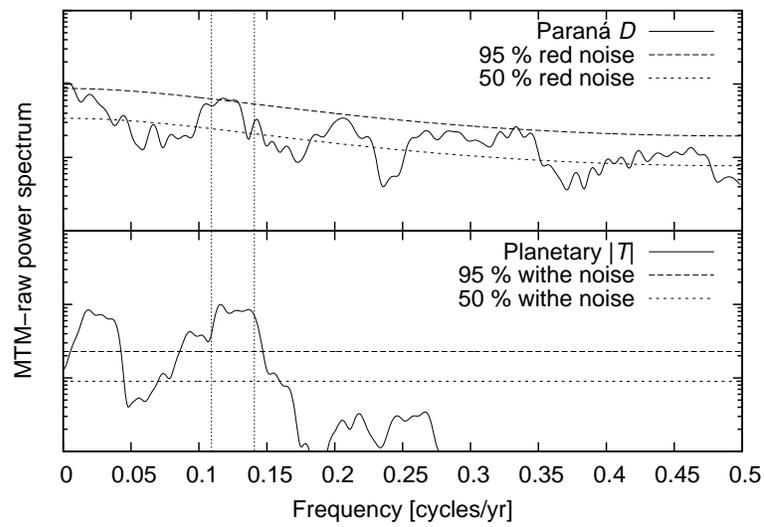}
\caption{MTM spectra comparison of Paran\'a $D$ series and  planetary $|T|$ torque. The vertical doted lines 
mark the common 7-9 yr band considered 
by \cite{AK11}. Inside this band, the spectral power of $D$ is significant at 50-95$\%$. The simple inspection 
of this figure shows that $D$ signal is significant at this levels in a broader sub-decadal band.}   
\end{figure}

\begin{figure}
\includegraphics[scale=1]{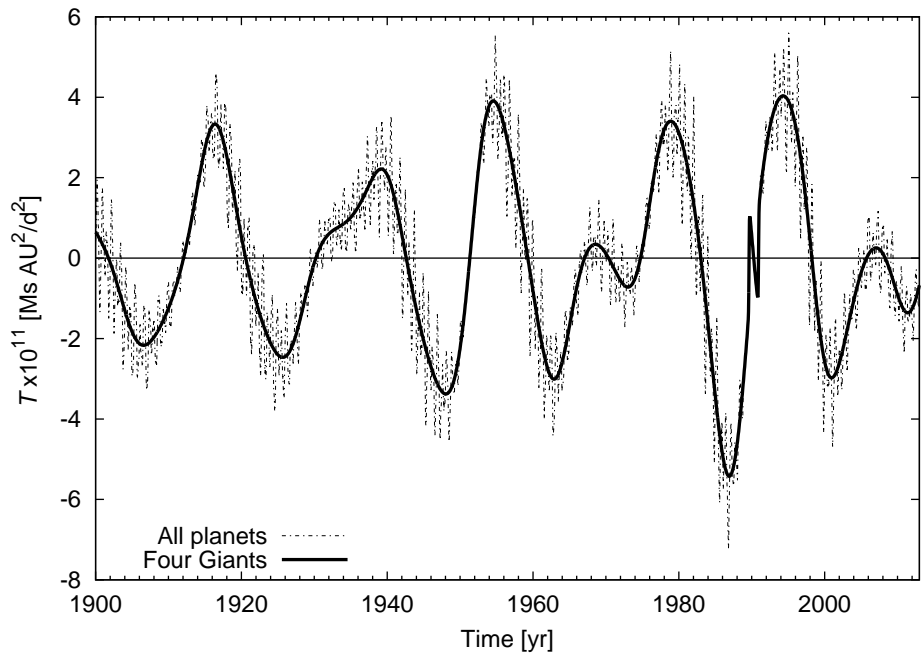}
\caption{Planetary $T$ torque from 1900 to 2013 A.D. Dashed line: all the planets included; solid line: only giant planets. 
Physical units: solar mass (Ms), astronomical unity (AU) and  days (d).}   
\end{figure}

\begin{figure}
\includegraphics[scale=1]{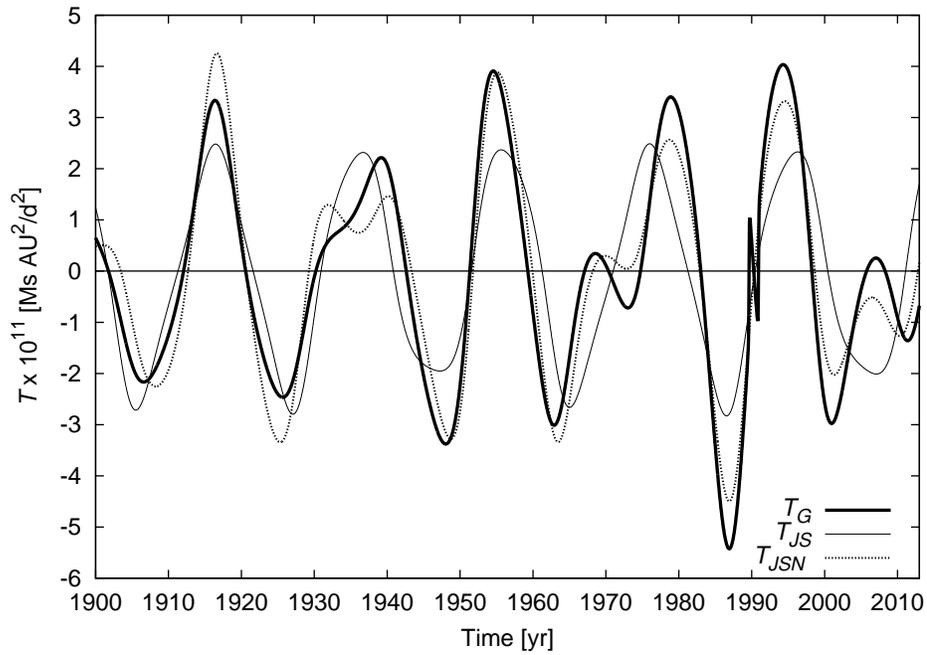}
\caption{Comparison between $T$ torque of only giant planets ($T_G$), $T$ from Jupiter and Saturn ($T_{JS}$), 
and from Jupiter, Saturn and Neptune ($T_{JSN}$) 
subsystem planets. Surpraisingly, the rapid variation (lesser than 1 yr) in $T$ signal around 1990 is due only 
to giant planets. The quasi-conjunction of Jupiter with other Giants forces to $T$ to be near zero. 
 The $T$ ``perturbation" around 1935 is due to the normal dynamics that strongly involve Neptune planet. }   
\end{figure}

\begin{figure}
\includegraphics[scale=1]{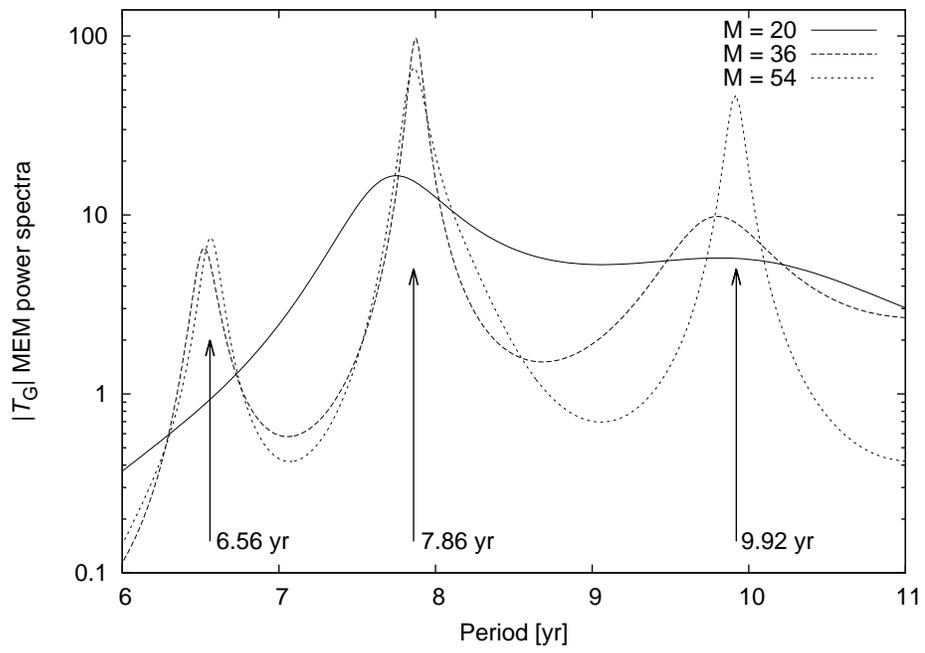}
\caption{Maximum entropy (MEM) spectra for $|T_G|$ with differents number of  poles $(M$) used. }   
\end{figure}

\begin{figure}
\includegraphics[scale=0.2]{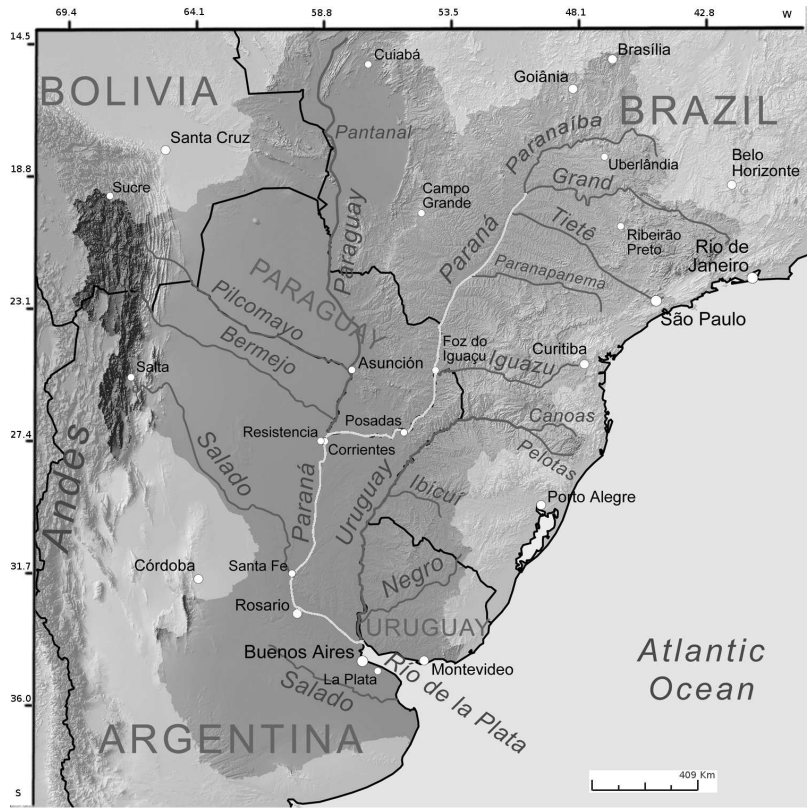}
\caption{Paran\'a's basin. The main rivers of Paran\'a system are showed. Corrientes city in Argentina is marked. 
Other South-American cities in Uruguay, Brazil, Bolivia and Paraguay are also indicated. }   
\end{figure}

\begin{figure}
\includegraphics[scale=1]{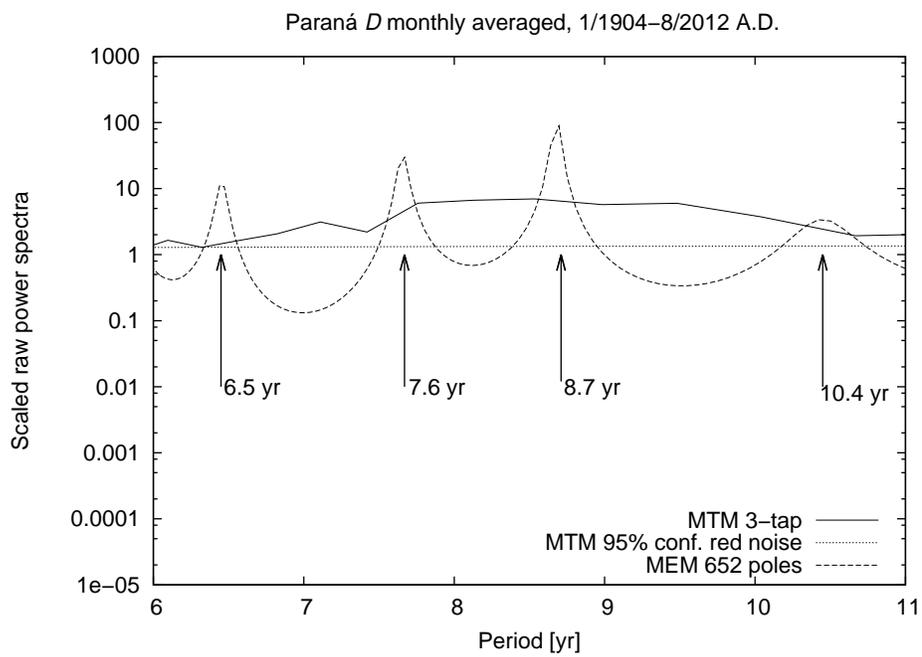}
\caption{MTM and MEM Spectra of Paran\'a's monthly $D$ series (1304 data values used). }   
\end{figure}

\begin{figure}
\includegraphics[scale=1.1]{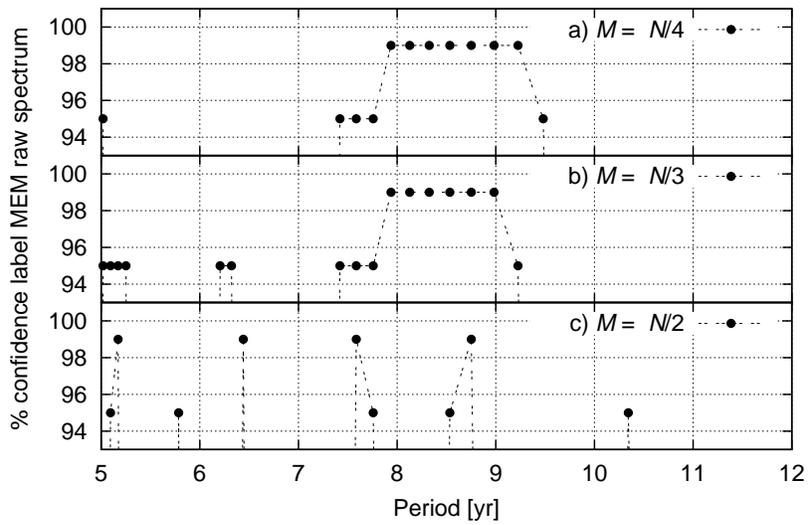}
\caption{Results of the Monte Carlo permutation test ({\tt MAXEMPER} software) performed to Paran\'a monthly $D$ series, showing the significance 
of these peaks against red noise null hypothesis. The pole order $M$ is showed (number of data $N$= 1304). }   
\end{figure}
    
\begin{figure}
\includegraphics[scale=1]{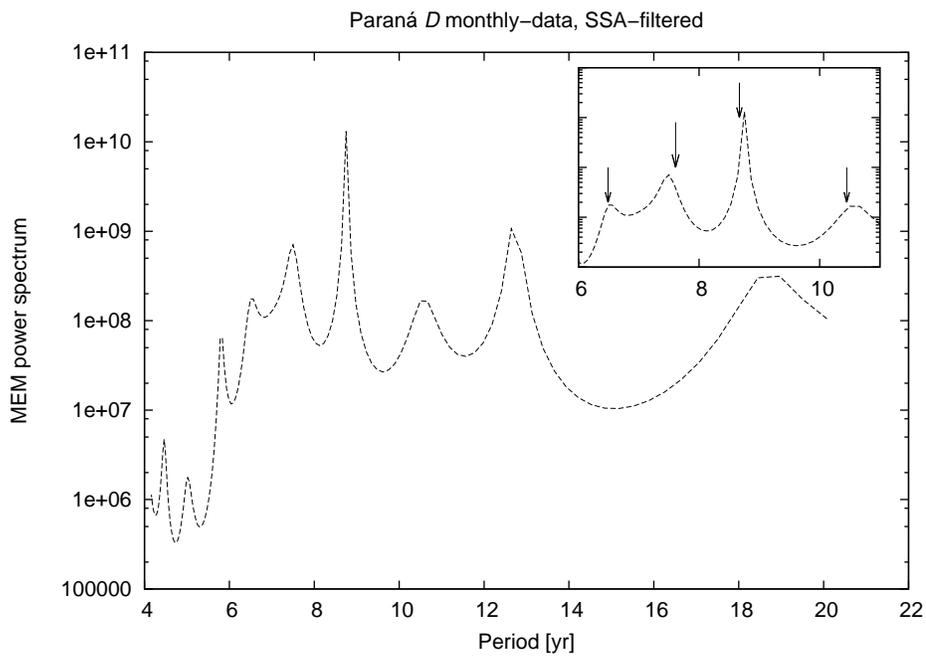}
\caption{MEM spectrum of the filtered $D$ monthly series showing the disappearance of noisy-peaks ($M = N/2$). 
The small window shows in detail that the same four peaks appear in the sub-decadal band. For comparison, the arrows indicate 
the peaks of the raw-annual data. }   
\end{figure}

\begin{figure}
\includegraphics[scale=1]{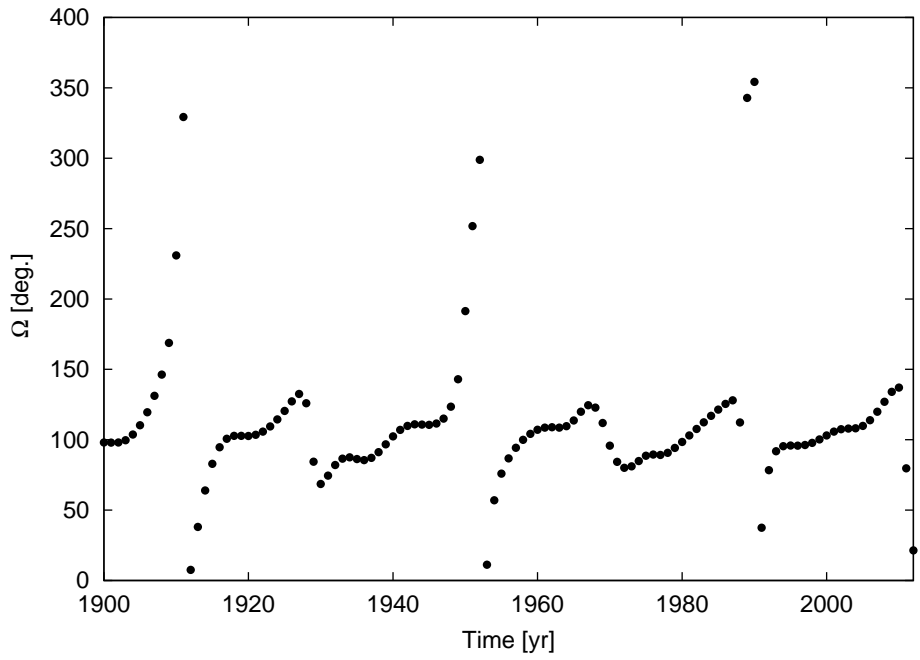}
\caption{Ascending node evolution of the solar barycentric orbit (one point per year), the same oscillations are performed
 by $\bf{L}$ vector around the $z$-axis of the inertial system, in this case, the ecliptical J2000.0 system. }   
\end{figure}

\begin{figure}
\includegraphics[scale=1.1]{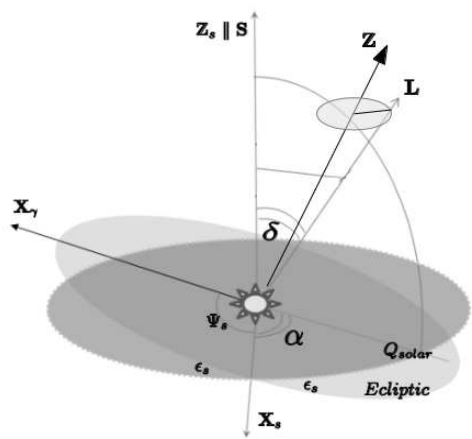}
\caption{Sketch of the IHS system showing the solar equator ($Qsolar$), the ecliptical plane (the $x$-$y$ plane of the inertial system) 
and its pole ${\bf Z}$ (the $z$-axis of the inertial system). The angles of orientation $\epsilon_s$ and $\Psi_s$ 
between the IHS and the inertial system are indicated (the orientation between ${\bf S}$ and ${\bf Z}$ was largely exagerated by convenience). 
The ${\bf L}$ vector has a precessional-like movement around $z$-inertial axis (it was idealized by the shadowed ellipse). 
The position angles of ${\bf L}$ respect to ${\bf S}$ ($\alpha$ and $\delta$) are indicated with double arcs. }   
\end{figure}

\begin{figure}
\includegraphics[scale=1.1]{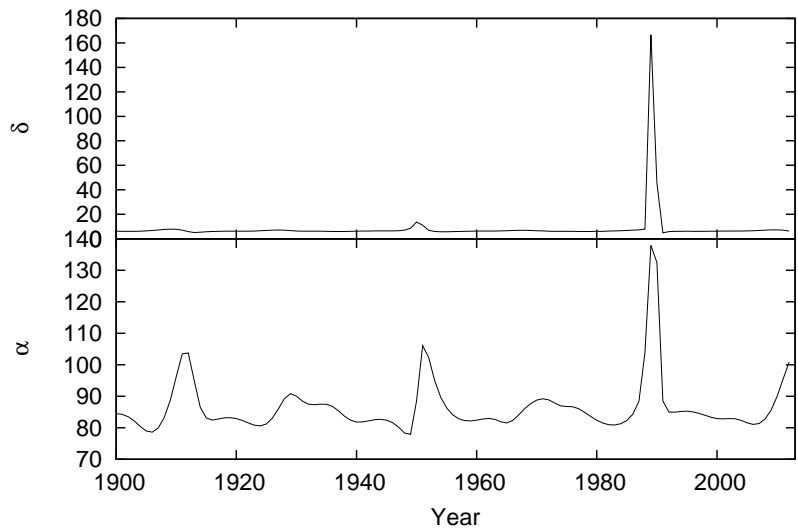}
\caption{$\alpha$ and $\delta$ evolution in the studied period. The most important peaks occur each $\sim$ 38 yr, due to Jupiter, Saturn 
and Neptune alignments. The greatest peak around 1990 involve the four giant planets. }   
\end{figure}

\begin{figure}
\includegraphics[scale=1.1]{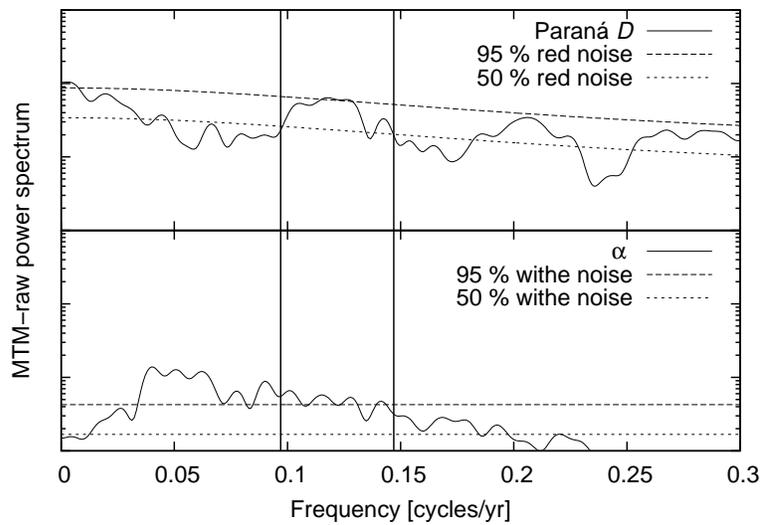}
\caption{MTM spectra of Paran\'a and $\alpha$ series. 
The vertical lines mark the band at which the spectral power of $D$ is significant at 50-95$\%$. 
Surprisingly, in this sub-decadal band four peaks seem to be very similar in both series. 
Certainly, Paran\'a spectrum  is more similar to $\alpha$ spectrum than $|T_G|$ spectrum.
Also, a prominent bi-decadal band is also seen in 
$\alpha$ spectrum. }   
\end{figure}

\begin{figure}
\includegraphics[scale=1]{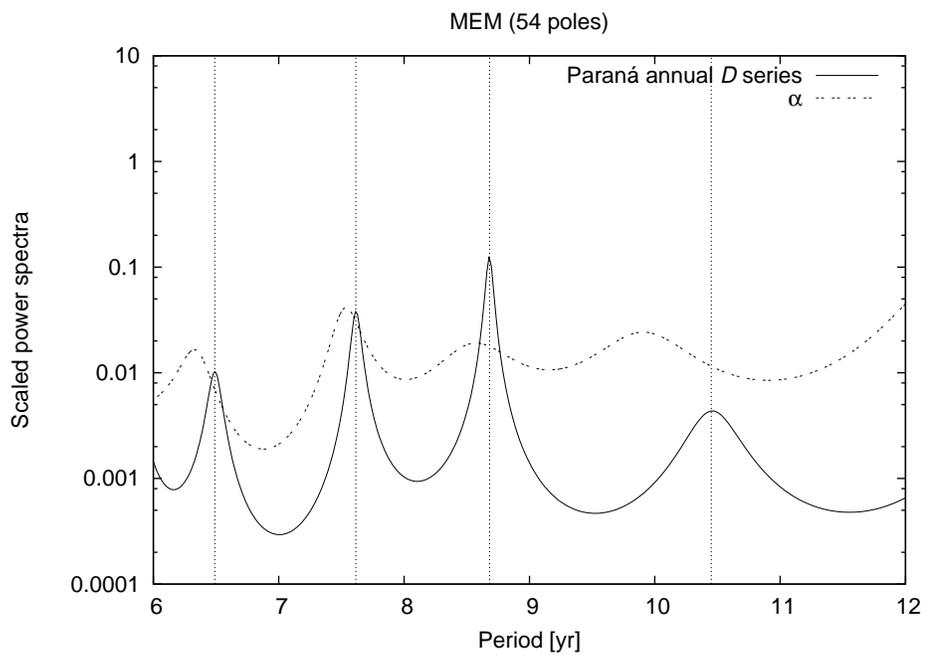}
\caption{MEM spectra of Paran\'a $D$ series and $\alpha$ series. It confirms the
 coincidence between four peaks. Therefore in the sub-decadal band Paran\'a's $D$ series has virtually 
 the same frequencies than $\alpha$ series (the porcentual error is lesser than 5$\%$). }   
\end{figure}

\begin{figure}
\includegraphics[scale=1]{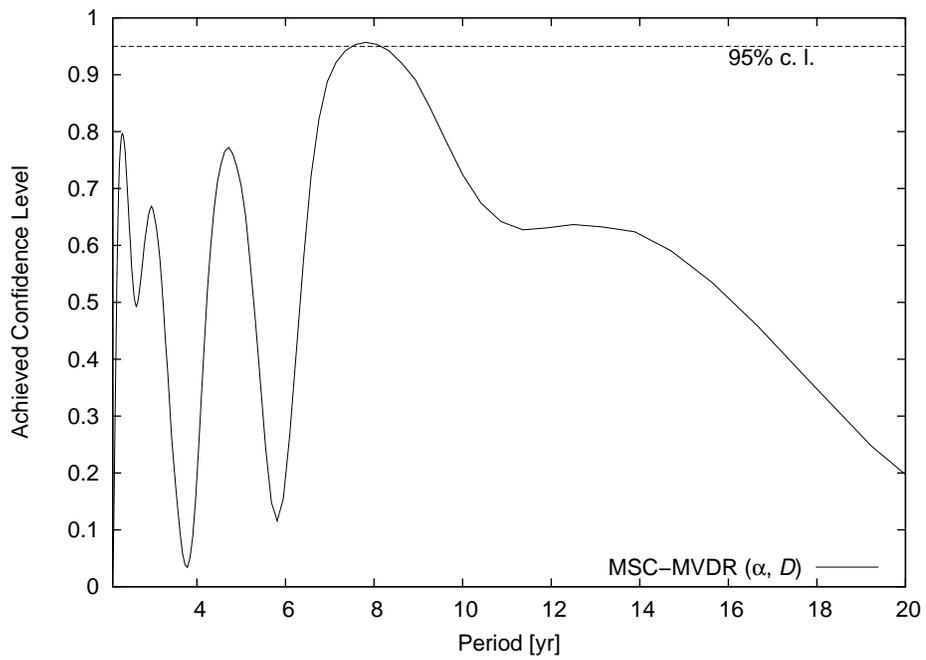}
\caption{Achieved Confidence Level (ACL) of the MSC (MVDR method), with Monte Carlo permutation test,
 between Paran\'a $D$ series and $\alpha$ series. Periodicities 
around 8 yr are statistically significant at 95$\%$ confidence level.}
\end{figure}

\begin{figure}
\includegraphics[scale=1]{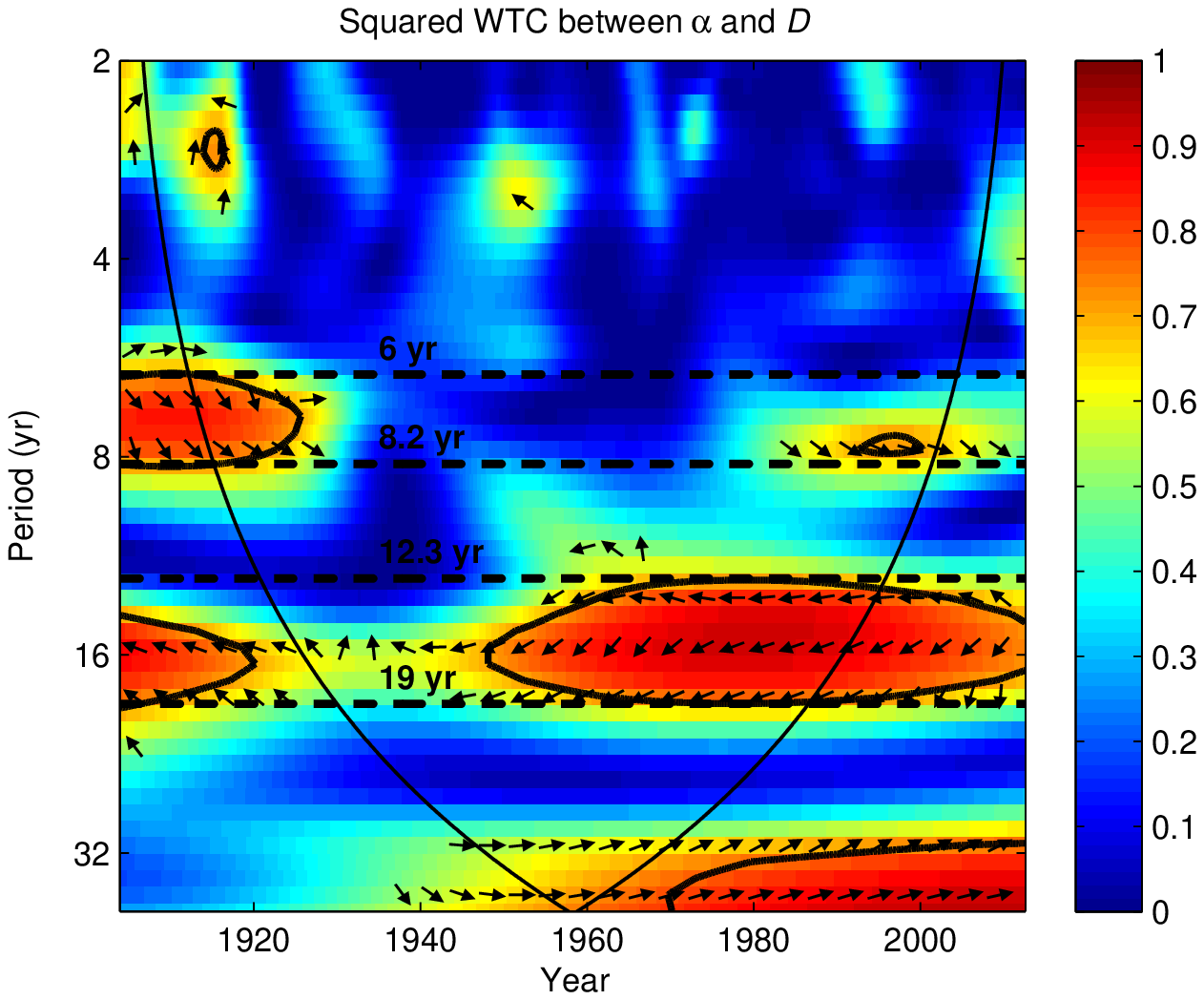}
\caption{Squared WTC coherence between $D$ and $\alpha$ series. It confirms the coherence's intermittency around 
8-yr band. The horizontal dashed lines are the approximated bounds of 95$\%$ confidence level zones, they   
marks periodicities between 6-8.2 yr and 12.3-19 yr.}
\end{figure}


\begin{thebibliography}{00}

\bibitem[Abreu et al.(2012)]{Abreuetal2012} Abreu, J.~A., Beer, J., Ferriz-Mas, A., 
McCracken, K.~G., Steinhilber, F.\ 2012.\ Is there a planetary influence on solar activity?\ Astronomy and Astrophysics 548, A88. 
\bibitem[Agnihotri et al.(2011)]{Agnihotrietal2011} Agnihotri, R., Dutta, 
K., Soon, W.\ 2011.\ Temporal derivative of Total Solar Irradiance and 
anomalous Indian summer monsoon: An empirical evidence for a Sun-climate 
connection.\ Journal of Atmospheric and Solar-Terrestrial Physics 73, 
1980-1987. 
\bibitem[Antico and Kr\"ohling(2011)]{AK11} Antico, A., 
Kr{\"o}hling, D.~M.\ 2011.\ Solar motion and discharge of Paran{\'a} River, 
South America: Evidence for a link.\ Geophysical Research Letters 38, 
19401. 
\bibitem[Benesty et al.(2006)]{Benesty} Benesty, J., Chen, J., Huang, Y.\ 2006. \ Estimation of the coherence 
function with the MVDR approach. Acoustics, Speech and Signal Processing. ICASSP 2006 Proceedings 3, 500-503. 
doi: 10.1109/ICASSP.2006.1660700.
\bibitem[Blizard(1981)]{Blizard1981} Blizard, J.~B.\ 1981.\ Solar 
Motion and Solar Activity.\ Bulletin of the American Astronomical Society 
13, 876. 
\bibitem[Brown(1900)]{Brown1900} Brown, E.~W.\ 1900.\ A possible 
explanation of the sun-spot period.\ Monthly Notices of the Royal 
Astronomical Society 60, 599. 
\bibitem[Burlaga(1984)]{Burlaga1984} Burlaga, L.~F.\ 1984.\ MHD 
processes in the outer heliosphere.\ Space Science Reviews 39, 255-316.
\bibitem[Cameron and Sch{\"u}ssler(2013)]{CameronandSchussler2013} Cameron, R.~H., Sch{\"u}ssler, 
M.\ 2013.\ No evidence for planetary influence on solar activity.\ Astronomy and Astrophysics 557, A83.
\bibitem[Cionco and Compagnucci(2012)]{CC12} Cionco, R.~G., 
Compagnucci, R.~H.\ 2012.\ Dynamical characterization of the last prolonged 
solar minima.\ Advances in Space Research 50, 1434-1444.
\bibitem[Chambers(1999)]{Chambers1999} Chambers, J.~E.\ 1999.\ A 
hybrid symplectic integrator that permits close encounters between massive 
bodies.\ Monthly Notices of the Royal Astronomical Society 304, 793-799.
\bibitem[Chang et al.(2012)]{Changetal2012} Chang, Y.-L., 
Bodenheimer, P.~H., Gu, P.-G.\ 2012.\ Coupled Evolutions of the Stellar 
Obliquity, Orbital Distance, and Planet's Radius due to the Ohmic 
Dissipation Induced in a Diamagnetic Hot Jupiter around a Magnetic T Tauri 
Star.\ The Astrophysical Journal 757, 118. 
\bibitem[Charv\'atov\'a(2009)]{Charvatova2009} Charv{\'a}tov{\'a}, 
I.\ 2009.\ Long-term predictive assessments of solar and geomagnetic 
activities made on the basis of the close similarity between the solar 
inertial motions in the intervals 1840 1905 and 1980 2045.\ New Astronomy 
14, 25-30. 
\bibitem[Charv\'atov\'a and St{\v r}e{\v s}t{\'{\i}}k(2004)]{CharvatovaandStrestik2004} 
Charv{\'a}tov{\'a}, I., St{\v 
r}e{\v s}t{\'{\i}}k, J.\ 2004.\ Periodicities between 6 and 16 years in 
surface air temperature in possible relation to solar inertial motion.\ 
Journal of Atmospheric and Solar-Terrestrial Physics 66, 219-227. 
\bibitem[Chang(2009)]{Chang09} Chang, H.-Y.  2009, Periodicity of 
North South asymmetry of sunspot area revisited: Cepstrum analysis, New 
Astronomy, 14, 133-138.
\bibitem[Cubasch et al.(2006)]{Cubaschetal2006} Cubasch, U., Zorita, 
E., Kaspar, F., Gonzalez-Rouco, J.~F., von Storch, H., Pr{\"o}mmel, K.\ 
2006.\ Simulation of the role of solar and orbital forcing on climate.\ 
Advances in Space Research 37, 1629-1634.
\bibitem[de Petris and Paquini(2007)]{dpp} Depetris P.J. y Pasquini A.I. 2007. 
The Geochemistry of the Paraná River: An 
Overview. In M.H. Iriondo, J.C. Paggi, y M.J. Parma (Eds.). The Middle 
Paraná River: Limnology of a Subtropical Wetland. Springer-Verlag Berlin 
Heidelberg.
\bibitem[Fairbridge and Shirley(1987)]{FairbridgeandShirley1987} Fairbridge, 
R.~W., Shirley, J.~H.\ 1987.\ Prolonged minima and the 179-yr cycle of the 
solar inertial motion.\ Solar Physics 110, 191-210.
\bibitem[Folkner et al.(2014)]{Folkneretal14} Folkner, W. M., Williams, J. G., Boggs, 
D. H., Park, R. S., Kuchynka, P.  2014, The Planetary and Lunar 
Ephemerides DE430 and DE431, Interplanetary Network Progress Report, 196, 
1.
\bibitem[Fr{\"a}nz and Harper(2002)]{FranzandHarper2002} Fr{\"a}nz, M., Harper, D.\ 2002.
\ Heliospheric coordinate systems.\ Planetary and Space Science 50, 217-233. 
\bibitem[Georgieva et al.(2012)]{Georgievaetal12} Georgieva, K., Kirov, 
 B., Kouck{\'a} Kn{\'{\i}}{\v z}ov{\'a}, P., Mo{\v s}na, Z., Kouba, D., 
 Asenovska, Y.\ 2012.\ Solar influences on atmospheric circulation.\ Journal 
 of Atmospheric and Solar-Terrestrial Physics 90, 15-25.
 \bibitem[Ghil et al.(2002)]{Ghiletal2002} Ghil, M., and 10 
 colleagues 2002.\ Advanced Spectral Methods for Climatic Time Series.\ 
 Reviews of Geophysics 40, 1003.
 \bibitem[Goldreich and Peale(1966)]{GP66} Goldreich, P., 
 Peale, S.\ 1966.\ Spin-orbit coupling in the solar system.\ The 
 Astronomical Journal 71, 425.
 \bibitem[Goldreich and Soter(1966)]{GoldreichandSoter1966} Goldreich, P., 
 Soter, S.\ 1966.\ Q in the Solar System.\ Icarus 5, 375-389. 
\bibitem[Gray et al.(2010)]{Grayetal2010} Gray, L.~J., and 14 
 colleagues 2010.\ Solar Influences on Climate.\ Reviews of Geophysics 48, 
 1-53.
\bibitem[Grinsted et al.(2004)]{Grinsted} Grinsted, A., Moore, J. C., Jevrejeva, S.  2004, Application of the 
cross wavelet transform and wavelet coherence to geophysical time series, 
Nonlinear Processes in Geophysics, 11, 561-566.
\bibitem[Holm(2014a)]{Holm2014a}Holm, S.  2014a, On the alleged coherence 
between the global temperature and the sun's movement, Journal of 
Atmospheric and Solar-Terrestrial Physics, 110, 23-27. 
\bibitem[Holm(2014b)]{Holm2014b} Holm, S.  2014b, Corrigendum to ``On the 
alleged coherence between the global temperature and the sun's movement'';, 
Journal of Atmospheric and Solar-Terrestrial Physics, 119, 230-231 
\bibitem[Humphreys(1910)]{H1910} Humphreys, W.~J.\ 1910.\ 
Solar Disturbances and Terrestrial Temperatures.\ The Astrophysical Journal 
32, 97-111.
\bibitem[Hung(2007)]{Hung2007} Hung C.-C., 2007. Apparent Relations Between Solar Activity and Solar Tides Caused by the Planets. NASA 
Technical Memorandum TM 2007-214817.
\bibitem[IPCC(2007)]{IPCC2007} IPCC: Climate Change 2007: The Physical Science Basis. Contribution of Working Group I to
 the Fourth Assessment Report of the Intergovernmental Panel on Climate Change [Solomon, S., D. Qin, M. Manning, Z. 
Chen, M. Marquis, K.B. Averyt, M.Tignor and H.L. Miller (eds.)]. Cambridge University Press, Cambridge, United 
Kingdom and New York, NY, USA.
\bibitem[Javaraiah(2015)]{Java15} Javaraiah, J.  2015, Long-term 
variations in the north-south asymmetry of solar activity and solar cycle 
prediction, III: Prediction for the amplitude of solar cycle 25, New 
Astronomy, 34, 54-64.
 \bibitem[Javaraiah(2005)]{Javaraiah2005} Javaraiah, J.\ 2005.\ Sun's 
 retrograde motion and violation of even-odd cycle rule in sunspot 
 activity.\ Monthly Notices of the Royal Astronomical Society 362, 
 1311-1318.
\bibitem[Jones et al.(2012)]{Jonesetal2012} Jones, G.~S., Lockwood, 
M., Stott, P.~A.\ 2012.\ What influence will future solar activity changes 
over the 21st century have on projected global near-surface temperature 
changes?\ Journal of Geophysical Research (Atmospheres) 117, D05103, 1-13.
\bibitem[Jose(1965)]{Jose1965} Jose, P.~D.\ 1965.\ Sun's motion 
and sunspots.\ The Astronomical Journal 70, 193. 
\bibitem[Juckett(2003)]{Juckett2003} Juckett, D. A.\ 2003.\ Temporal variations of low-order spherical harmonic
 representations of sunspot group patterns: Evidence for solar spin-orbit coupling.\ Astronomy and Astrophysics 399, 731-741. 
\bibitem[Juckett(2000)]{Juckett2000} Juckett, D. A.\ 2000.\ Solar 
activity cycles, north/south asymmetries, and differential rotation 
associated with solar spin-orbit variations.\ Solar Physics 191, 201-226. 
\bibitem[Krepper et al.(2008)]{Krepperetal2008} Krepper, C. M., García N., Jones P. \ 2008\, Low‐frequency
response of the upper Paran\'a basin, Int. J. Climatol. 28, 351–360.
\bibitem[Landscheidt(2000)]{Landscheidt2000} Landscheidt, T. \ 2000.\ 
River Po discharges and cycles of solar activity. Hydrol. Sci. J. 45(3), 491–493
\bibitem[Landscheidt(1999)]{Landscheidt1999} Landscheidt, T.\ 1999.\ 
Extrema in sunspot cycle linked to Sun's motion..\ Solar Physics 189, 
415-426.
\bibitem[Landscheidt(1987)]{Landscheidt1987} Landscheidt, T.\ 1987.\ 
Cyclic distribution of energetic X-ray flares.\ Solar Physics 107, 195-199.
\bibitem[Leal-Silva and Velasco Herrera(2012)]{Leal-SilvaandVelascoHerrera2012} 
Leal-Silva, M.~C., Velasco Herrera, V.~M.\ 2012.\ Solar forcing on the ice 
winter severity index in the western Baltic region.\ Journal of Atmospheric 
and Solar-Terrestrial Physics 89, 98-109.
\bibitem[Makarov et al.(1997)]{Makarovetal} Makarov, V.~I., Tlatov, 
A.~G., Callebaut, D.~K.\ 1997.\ Long-Term Variations of the Torsional 
Oscillations of the Sun.\ Solar Physics 170, 373-388. 
\bibitem[Mokhov and Smirnov(2006)]{MS} Mokhov, I.~I., 
Smirnov, D.~A.\ 2006.\ El Ni{\~n}o-Southern Oscillation drives North 
Atlantic Oscillation as revealed with nonlinear techniques from climatic 
indices.\ Geophysical Research Letters 33, 3708. 
\bibitem[Murray and Dermott(1999)]{MurrayandDermott1999} Murray, C.~D., 
Dermott, S.~F.\ 1999.\ Solar system dynamics.\ Solar system dynamics by 
Murray, C.~D., 1999 . 
\bibitem[Okal and Anderson(1975)]{OkalandAnderson1975} Okal, E., Anderson, 
D.~L.\ 1975.\ On the planetary theory of sunspots.\ Nature 253, 511-513. 
\bibitem[Palu{\v s} and Novotn{\'a}(2009)]{Palus&Novotna} Palu{\v s}, 
M., Novotn{\'a}, D.\ 2009.\ Phase-coherent oscillatory modes in solar and 
geomagnetic activity and climate variability.\ Journal of Atmospheric and 
Solar-Terrestrial Physics 71, 923-930. 
\bibitem[Pardo-Ig{\'u}zquiza and Rodr{\'{\i}}guez-Tovar(2005)]{PIRT05} 
Pardo-Ig{\'u}zquiza, E., 
Rodr{\'{\i}}guez-Tovar, F.~J.\ 2005.\ MAXENPER: a program for maximum 
entropy spectral estimation with assessment of statistical significance by 
the permutation test.\ Computers and Geosciences 31, 555-567. 
\bibitem[Peale(2005)]{Peale2005} Peale, S.~J.\ 2005.\ The free 
precession and libration of Mercury.\ Icarus 178, 4-18. 
\bibitem[Peale and Gold(1965)]{PealeandGold1965} Peale, S.~J., Gold, T.\ 
1965.\ Rotation of the Planet Mercury.\ Nature 206, 1240-1241. 
\bibitem[Penland et al.(1991)]{Penlandetal1991} Penland, C., Ghil, M., 
Weickmann, K.~M.\ 1991.\ Adaptive filtering and maximum entropy spectra 
with application to changes in atmospheric angular momentum.\ Journal of 
Geophysical Research 96, 22659-22671.
\bibitem[Perryman and Schulze-Hartung(2011)]{PMAC11} Perryman, M.~A.~C., Schulze-Hartung, T.\ 2011.\ 
The barycentric motion of exoplanet host stars. Tests of solar spin-orbit coupling.\ Astronomy and Astrophysics 525, A65. 
\bibitem[Pinto et al.(1992)]{Pintoetal91} Pinto, O., Jr., Gonzalez, 
W.~D., Pinto, I.~R.~C., Gonzalez, A.~L.~C., Mendes, O., Jr.\ 1992.\ The 
South Atlantic Magnetic Anomaly - Three decades of research.\ Journal of 
Atmospheric and Terrestrial Physics 54, 1129-1134. 
\bibitem[Pinto Neto et al.(2013)]{Pintoetal13} Pinto Neto, O., 
Pinto, I.~R.~C.~A., Pinto, O.\ 2013.\ The relationship between thunderstorm 
and solar activity for Brazil from 1951 to 2009.\ Journal of Atmospheric 
and Solar-Terrestrial Physics 98, 12-21. 
\bibitem[Poincar{\'e}(1910)]{Poincare} Poincar{\'e}, H.\ 1910.\ 
Sur la pr{\'e}cession des corps d{\'e}formables.\ Bulletin Astronomique, 
Serie I 27, 321-356.
\bibitem[Poulianovsky and Usoskin(2014)]{PoulianovskyUsoskin} Poluianov, S., Usoskin, I.  2014, Critical Analysis of a Hypothesis 
of the Planetary Tidal Influence on Solar Activity, Solar Physics, 289, 
2333-2342.
\bibitem[Press et al.(1992)]{Pressetal1992} Press, W.~H., Teukolsky, 
S.~A., Vetterling, W.~T., Flannery, B.~P.\ 1992.\ Numerical recipes in 
FORTRAN. The art of scientific computing.\ Cambridge: University Press, 
|c1992, 2nd ed.
\bibitem[Robertson and  Mechoso(1998)]{RM98}
Robertson, A. W., Mechoso C. R. 1998, Interannual and decadal cycles
in river flows of southeastern South America, J. Clim., 11, 2570–2581.
\bibitem[Rozelot(1994)]{Rozelot1994} Rozelot, J.~P.\ 1994.\ On the 
stability of the 11-year solar cycle period (and a few others).\ Solar 
Physics 149, 149-154. 
\bibitem[Scafetta(2014)]{Scafetta2014}Scafetta, N. 2014, Discussion on the spectral coherence between planetary, 
solar and climate oscillations: a reply to some critiques, Astrophysics and 
Space Science, in press, DOI: 10.1007/s10509-014-2111-8.
\bibitem[Scafetta(2012a)]{Scafetta2012a} Scafetta, N.\ 2012a.\ Does the 
Sun work as a nuclear fusion amplifier of planetary tidal forcing? A 
proposal for a physical mechanism based on the mass-luminosity relation.\ 
Journal of Atmospheric and Solar-Terrestrial Physics 81, 27-40. 
\bibitem[Scafetta(2012b)]{Scafetta2012b} Scafetta, N.\ 2012b.\ 
Multi-scale harmonic model for solar and climate cyclical variation 
throughout the Holocene based on Jupiter-Saturn tidal frequencies plus the 
11-year solar dynamo cycle.\ Journal of Atmospheric and Solar-Terrestrial 
Physics 80, 296-311.
\bibitem[Scafetta(2010)]{Scafetta2010} Scafetta, N.\ 2010.\ 
Empirical evidence for a celestial origin of the climate oscillations and 
its implications.\ Journal of Atmospheric and Solar-Terrestrial Physics 72, 
951-970. 
\bibitem[Scafetta and West(2007)]{ScafettaandWest2007} Scafetta, N., West, B.~J.\ 2007.\ Phenomenological 
reconstructions of the solar signature in 
the Northern Hemisphere surface temperature records since 1600.\ Journal of 
Geophysical Research (Atmospheres) 112, 24. 
\bibitem[Scafetta and Willson(2013a)]{ScafettaandWillson2013a} Scafetta, N., Willson, R.~C.\ 2013a.\ 
Empirical evidences for a planetary modulation of total solar irradiance and the 
TSI signature of the 1.09-year Earth-Jupiter conjunction cycle.\ Astrophysics and Space Science 287. 
\bibitem[Scafetta and Willson(2013b)]{ScafettaandWillson2013b} Scafetta, N., Willson, R.~C.\ 2013b.\ Planetary harmonics
in the historical Hungarian aurora record (1523-1960).\ Planetary and Space Science 78, 38-44. 
\bibitem[Shirley(2014)]{Shirley2014} Shirley, J. M. 2014, Solar system dynamics and global-scale dust storms 
on Mars. Icarus, in press, DOI: http://dx.doi.org/10.1016/j.icarus.2014.09.038.
\bibitem[Shirley(2009)]{Shirley2009} Shirley, J.~H.\ 2009.\ Have We 
Entered a 21st Century Prolonged Minimum of Solar Activity? Updated 
Implications of a 1987 Prediction.\ AAS/Solar Physics Division Meeting \#40 
40, \#11.08.
\bibitem[Shirley(2006)]{Shirley2006} Shirley, J.~H.\ 2006.\ Axial 
rotation, orbital revolution and solar spin-orbit coupling.\ Monthly 
Notices of the Royal Astronomical Society 368, 280-282. 
\bibitem[Soon et al.(2011)]{Soonetal2011} Soon, W., Dutta, K., 
Legates, D.~R., Velasco, V., Zhang, W.\ 2011.\ Variation in surface air 
temperature of China during the 20th century.\ Journal of Atmospheric and 
Solar-Terrestrial Physics 73, 2331-2344. 
\bibitem[Stott et al.(2003)]{Stottetal2003} Stott, P.~A., Jones, 
G.~S., Mitchell, J.~F.~B.\ 2003.\ Do Models Underestimate the Solar 
Contribution to Recent Climate Change?\ Journal of Climate 16, 4079-4093.
\bibitem[Tan and Cheng(2013)]{TanandCheng2013} Tan, B., Cheng, Z.\ 2013.\ The 
mid-term and long-term solar quasi-periodic cycles and the possible relationship with 
planetary motions.\ Astrophysics and Space Science 343, 511-521. 
\bibitem[Tomasino et al.(2000)]{Tomasinoetal2000} Tomasino, M., Dalla Valle, F.  2000.\, 
Natural climatic changes and solar cycles: An analysis of hydrological time series, Hydrol. Sci. J., 45(3),
477 – 490.
\bibitem[Wolf(1859)]{Wolf1859} Wolf, R.\ 1859.\ Extract of a 
Letter to Mr. Carrington.\ Monthly Notices of the Royal Astronomical 
Society 19, 85-86.
\bibitem[Wolff and Patrone(2010)]{WP10} Wolff, C.~L., 
Patrone, P.~N.\ 2010.\ A New Way that Planets Can Affect the Sun.\ Solar 
Physics 266, 227-246. 
\bibitem[Wood and Wood(1965)]{WoodandWood1965} Wood, R.~M., Wood, K.,\ 1965.\ Solar Motion and Sunspot Comparison.\ Nature 208, 129-131. 
\bibitem[Zanchettin et al.(2008)]{Zanchettinetal2008} Zanchettin, D., 
Rubino, A., Traverso, P., Tomasino, M.\ 2008.\ Impact of variations in 
solar activity on hydrological decadal patterns in northern Italy.\ Journal 
of Geophysical Research (Atmospheres) 113, 12102.
\bibitem[Zaqarashvili(1997)]{Zava} Zaqarashvili, T.~V.\ 
1997.\ On a Possible Generation Mechanism for the Solar Cycle.\ The 
Astrophysical Journal 487, 930.


 \end{thebibliography}
\end{document}